\documentclass[11pt,a4paper]{article}
\pdfoutput=1

\usepackage{jheplike}
\usepackage{amsmath,amsfonts,amssymb}
\usepackage{array}
\usepackage{subfig}

\newcommand{\be}{\begin{equation}}
\newcommand{\ee}{\end{equation}}

\def\nn{\nonumber}
\def\pd{\partial}

\def\ub{\bar{u}}
\def\zb{\bar{z}}
\def\O{\mathcal{O}}

\def\K{\mathcal{K}}

\title{Tuning and Backreaction in F-term Axion Monodromy Inflation}

\author[a]{Arthur Hebecker,\note[a]
{\href{mailto:A.Hebecker@ThPhys.Uni-Heidelberg.de}
{A.Hebecker@ThPhys.Uni-Heidelberg.de}}}
\author[b]{Patrick Mangat,\note[b]
{\href{mailto:P.Mangat@ThPhys.Uni-Heidelberg.de}
{P.Mangat@ThPhys.Uni-Heidelberg.de}}}
\author[c]{Fabrizio Rompineve\note[c]
{\href{mailto:F.Rompineve@ThPhys.Uni-Heidelberg.de}
{F.Rompineve@ThPhys.Uni-Heidelberg.de}}}
\author[d]{and Lukas T.~Witkowski\note[d]
{\href{mailto:L.Witkowski@ThPhys.Uni-Heidelberg.de}
{L.Witkowski@ThPhys.Uni-Heidelberg.de}}}

\affiliation{Institute for Theoretical Physics, University of Heidelberg, \\ 
Philosophenweg 19, 69120 Heidelberg, Germany \vspace{0.1cm}}

\abstract{We continue the development of axion monodromy inflation, focussing in particular on the backreaction of complex structure moduli. In our setting, the shift symmetry comes from a partial large complex structure limit of the underlying type IIB orientifold or F-theory fourfold. The coefficient of the inflaton term in the superpotential has to be tuned small to avoid conflict with K\"ahler moduli stabilisation. To allow such a tuning, this coefficient necessarily depends on further complex structure moduli. At large values of the inflaton field, these moduli are then in danger of backreacting too strongly. To avoid this, further tunings are necessary. In weakly coupled type IIB theory at the orientifold point, implementing these tunings appears to be difficult if not impossible. However, fourfolds or models with mobile D7-branes provide enough structural freedom. We calculate the resulting inflaton potential and study the feasibility of the overall tuning given the limited freedom of the flux landscape. Our preliminary investigations suggest that, even imposing all tuning conditions, the remaining choice of flux vacua can still be large enough for such models to provide a promising path to large-field inflation in string theory.}

\begin{document}
\maketitle

\section{Introduction}
Realising large field inflation in string theory is notoriously difficult.  Suggested mechanisms include axion monodromy \cite{08033085,08080706}, models with axion alignment \cite{0409138}, and N-flation \cite{0507205, 07103883, 09031481, 11010048, 14012579} (see also \citep{9804177}).\footnote{For a field-theoretic implementation of axion monodromy inflation see \cite{08111989, 11010026}.} This area of research has received renewed interest \cite{14037507, 14043040, 14043542, 14043711, 14044268, 14045235, 14047127, 14047773, 14047496, 14047852, 14051685, 14052325, 14053652, 14057044, 14060341, 14070253, 14097075, 14100016, 14107522} (for a recent review see \cite{14095350}) due to the possible tensor mode observation by the BICEP2 experiment \cite{BICEP2}. While the interpretation of BICEP2 data in terms of tensor modes now appears less straightforward due to the considerable dust background detected by Planck \cite{14095738}, future combined analyses or even new measurements may still force us to focus on large field models. Also, independently of the data, we are attracted by the purely theoretical challenge of realising large field inflation in string theory.

In the present paper, we intend to face this challenge in the context of the type IIB/F-theory flux landscape. We focus on a recently proposed class of string-theoretic supergravity realisations of axion monodromy \cite{14043040, 14043542, 14043711}. The basic underlying idea of all these constructions is as follows: One considers settings where at least one modulus enjoys a shift-symmetric K\"ahler potential. This shift symmetry as well as the related periodicity of the axionic part of the modulus is then weakly broken by the superpotential, e.g.~due to an appropriate flux choice. This gives rise to an enlarged axion field range with a slowly rising potential, suitable for large-field inflation. 

As usual in string-theory inflation, moduli stabilisation is a critical issue. This problem was analysed in some detail in \cite{14043711} concerning K\"ahler moduli while, concerning complex structure moduli, high-scale flux-stabilisation was assumed in a somewhat simple-minded way. It is our primary intention to improve on this part of the analysis.

To be more specific, the central idea of \cite{14043711} (see also \cite{14053652, 14097075, 14107522}) was to use the shift symmetry of complex structure moduli (or, equivalently, D7-brane moduli) of the F-theory 4-fold in the large-complex-structure limit. For one of these moduli, which we denote by $u$, it was then assumed that its coefficients in the superpotential are small due to a tuning of the values of the other moduli through flux choice. To be specific, for a superpotential of the form
\be
W(z,u) = w(z) + a(z) u + \frac{b(z)}{2} u^2 + \ldots
\ee
one assumes that the coefficients $a(z)$, $b(z)$ etc.~are tuned small. Here $z$ denotes the set of all complex structure moduli different from $u$. 

One can consider situations where these superpotential coefficients do not depend on the other moduli at all. This has very recently been implemented in \cite{14097075}. However, due to the integrality of flux numbers the relevant coefficients then cannot be parametrically small. While other complex structure moduli can be made parametrically much heavier due to large flux numbers, lowering the inflaton mass to the phenomenologically required value or even just below the scale at which the notoriously light K\"ahler moduli are stabilised remains challenging.\footnote{Problems regarding the backreaction of K\" ahler moduli in large field inflation via shift-symmetric complex moduli have also been recently discussed in \cite{14107522}. There, K\" ahler moduli are stabilised in a racetrack scenario.}

Our strategy is somewhat different, following more closely the idea originally suggested in \cite{14043711}: We want to make the crucial coefficient of the inflaton field in the superpotential small by a standard landscape-type tuning \cite{0004134, 0602072}. In other words, we make use of the fact that this coefficient is the sum of many terms, each depending on several other moduli. The vacuum values of these moduli in turn depend on a high-dimensional integral flux vector. Due to a fine cancellation between the various terms, the value of the coefficient and hence the inflaton mass can then be made extremely small. However, there is a price to be paid: By the very definition of our approach the coefficient depends on other moduli. Thus, to ensure that these are not destabilised, all derivatives of this coefficient with respect to the moduli entering it must be tuned small. We explain how the resulting, highly tuned inflaton scalar potential can be derived from the general supergravity formula. Furthermore, following closely the strategy of \cite{0404116}, we estimate the required tuning and quantify under which conditions a tuning of this strength can be realised in a Calabi-Yau orientifold with a certain D3 tadpole and a certain number of cycles. 

In this work we will discuss 3-folds with orientifold projection, as well as F-theory 4-folds. In contrast to \cite{14043711, 14097075}, we only require one of the complex structure moduli to be in the large complex structure regime. This is sufficient to suppress the relevant set of instantons on the mirror 3/4-fold and hence to ensure the decisive leading-order shift-symmetric structure of the model. Such a ``partial large complex structure limit'' may be essential to avoid a potentially enormous fine-tuning price of being near the large complex complex structure point in moduli space \cite{0404116} (see however \cite{14050283}). The requirement of being in the physical domain of the moduli space will in general force some moduli in addition to $u$ to be in the large complex structure regime. However, this is still better than demanding the large complex structure limit for all moduli from the beginning. We find that the required tunings cannot be implemented in the case of 3-folds if the string coupling is to remain in a perturbative regime and if we do want to avoid destabilisation of the saxion partner of the inflaton. On the contrary, we observe that tuning the relevant coefficients in the superpotential is in principle possible for 4-folds.

While our overall conclusion is positive, models of the class we consider are highly non-generic or tuned. This appears to have a clear structural reason: If we want the coefficient of the inflaton to be parametrically small, it can not be a simple number -- it must depend on other moduli. Thus, when the inflaton moves over a large field range, these other moduli are in danger of being destabilised. This has to be prevented by further tunings. While we expect that this problem will also affect the proposals of \cite{14043040}, where the crucial superpotential coefficients are small due to the choice of a particular geometric regime (i.e.~again a moduli choice), the proposals in that paper are not sufficiently explicit to directly apply our considerations of moduli stabilisation to it. It will be interesting to go systematically through the classes of suggested large-field models and see which constructions can work with the least tuning, but this is beyond the scope of our paper. Notably, since the observational verdict concerning large or small field models is still out, one has the option of deciding that large field models are more tuned than certain (potentially non-tuned) small field constructions (see \cite{14042601} for a review of inflation models in string theory) and thus to {\it predict} a small tensor-to-scalar ratio from string theory, as attempted in \cite{12064034, 13033224}.

The paper is organised as follows. In section \ref{ch:tuningandbr} we examine the necessary tunings as well as backreaction analytically. In particular, in section \ref{sec:theproblem} we outline that a tuning of the coefficients $a(z)$ etc.~alone is not enough to arrive at a sufficiently flat potential for the inflaton field, and that the derivatives $\pd_z a(z)$ etc.~have to be small as well. In section \ref{sec:nogo3} we show that these tunings cannot be implemented in type IIB orientifolds, if the string coupling is to remain in a perturbative regime and if we do want to avoid destabilisation of the saxion partner of the inflaton. In contrast, models of axion monodromy with the desired properties can be successfully implemented in F-theory 4-folds, which we describe in section \ref{sec:4foldsetup}. In sections \ref{sec:branalytical}--\ref{sec:kaehlermoduli} we then study backreaction of complex structure moduli and the resulting effective inflaton potential analytically. Numerical examples are shown in section \ref{sec:brnum}. Last, we estimate the number of string vacua with the desired properties in section \ref{sec:landscape}. 


\section{Tuning and backreaction}
\label{ch:tuningandbr}
\subsection{The problems of tuning and backreaction}
\label{sec:theproblem}
Here we will briefly outline problems with tuning and backreaction in models of $F$-term axion monodromy inflation. We begin by collecting the necessary ingredients for such a model and review the philosophy. Axion monodromy setups require a shift-symmetric K\"ahler potential as well as a superpotential which breaks this shift symmetry. For the moment we consider
\be
\label{eq:themodel}
\mathcal{K}_{cs}=\mathcal{K}_{cs}(z, \zb, u+\bar{u}) \ , \qquad W=w(z) + a(z) u \ .
\ee
Here, $z$ stands collectively for a set of moduli $\{ z^i \}$. The K\"ahler potential is invariant under shifts $u \rightarrow u + i \alpha$. In our setting, the shift symmetry will arise from a partial large complex structure limit of the underlying type IIB orientifold or F-theory fourfold. If the superpotential was also invariant under this shift of $u$, the direction $y \equiv \textrm{Im}(u)$ would be exactly flat. It is this shift-symmetric direction $y$ which is identified as an inflaton candidate.\footnote{Please note the change of notation compared to \cite{14043711}, which discusses a similar inflation model in supergravity. There $\mathcal{K}_{cs}=\mathcal{K}_{cs}(z, \zb, c-\bar{c})$ with $c$ a D7-brane (or fourfold complex structure) modulus. The most important difference is that the inflaton in \cite{14043711} is given by $\textrm{Re}(c)$, while now we use $\textrm{Im}(u)$.} By including the term $a(z) u$ in $W$ the shift symmetry is broken and a potential for the inflaton is generated. 

By breaking the shift symmetry weakly one aims to keep the inflaton potential sufficiently flat for inflation to work and not to interfere with moduli stabilisation.\footnote{Alternatively, one could try to go to the regime $w(z) \gg 1$ \cite{14097075}. We compare the two different approaches in appendix \ref{sec:comparison}.} As the breaking is determined by the parameter $a(z)$ one expects the inflaton potential to be controllably flat by choosing this parameter small enough at the SUSY locus $z=z_{\star}$. In the following we will argue that this is not sufficient: in particular, we find that there are further parameters in the model which need to be tuned small.

We identify the $z$ and $u$ as complex structure moduli (or D7-brane position moduli) in a type IIB orientifold setting, or as F-theory fourfold complex structure moduli. In the threefold case we also include the axio-dilaton in the set of moduli labelled $z$. The potential responsible for moduli stabilisation as well as inflation is the supergravity scalar potential
\be
V =e^{\mathcal{K}} (\mathcal{K}^{I \bar{J}} D_I W \overline{D_J W} + \mathcal{K}^{T_{\gamma} \bar{T}_{\delta}} D_{T_{\gamma}} W \overline{D_{T_{\delta}} W} -3 |W|^2) \ ,
\ee
where $\mathcal{K} = - 2 \ln \mathcal{V} + \mathcal{K}_{cs}(z, \zb, u + \ub)$. The index $I$ runs over all moduli $z$ as well as $u$. We wish to embed our inflation model in setups where K\"ahler moduli are stabilised according to the Large Volume Scenario \cite{0502058}. In this case the last two terms cancel at leading order due to the no-scale structure in the K\"ahler moduli sector and we are left with\footnote{Indeed, ignoring $\alpha^{'}$- and instanton corrections coming from the blow-up cycles of a swiss-cheese CY threefold, the quadratic form containing the K\"ahler moduli approximately cancels with $-3\left|W\right|$. Then, the LVS K\"ahler moduli stabilisation can proceed in the well known way, giving rise to an AdS minimum with $V_{LVS} \sim - |W|^2/\mathcal{V}^3$. For the moment we ignore this extra contribution to \eqref{eq:SUGRAV01}. We will return to this when commenting on K\"ahler moduli stabilisation in section \ref{sec:kaehlermoduli}. }
\be
\label{eq:SUGRAV01} V = e^{\mathcal{K}} (\mathcal{K}^{I \bar{J}} D_I W \overline{D_J W}) \ .
\ee
Most importantly, we consider all fields as dynamical, i.e.~we do not integrate out all $z$ at this stage. To simplify the argument, we continue our analysis for only two fields, labelled by $z$ and $u$. The two $F$-terms entering \eqref{eq:SUGRAV01} are then given by
\begin{align}
D_u W \ &= D_u w + a + \mathcal{K}_u a u \ , \\
D_z W \ &= D_z w + (\pd_z a + \mathcal{K}_z a) u \ .
\end{align}
The values of $u$ and $z$ at the minimum of the F-term potential are obtained by solving the equations
\begin{equation}
D_{u}W=0, \quad D_{z}W=0.
\end{equation}
The latter can be interpreted as conditions on the derivatives of the K\" ahler potential at the minimum:
\begin{align}
\label{eq:dmin}
& D_{u}W=0\Rightarrow \mathcal{K}_{u}|_{min}=-\frac{a}{w+au}\Big|_{min}\\
& D_{z}W=0\Rightarrow \mathcal{K}_{z}|_{min}=-\frac{\partial_{z}w+\partial_{z}a\cdot u}{w+au}\Big|_{min}.
\end{align}
The inflaton potential will get contributions from both $F$-terms and takes the form:
\begin{equation}
\label{eq:fpotential1}
V=e^{\mathcal{K}}\Big[\mathcal{K}^{u\bar{u}}|\mathcal{K}_{u}a|^{2}+K^{z\bar{z}}|\partial_{z}a+\mathcal{K}_{z}a|^{2}+\mathcal{K}^{u\bar{z}}(\mathcal{K}_{u}a)(\overline{\partial_{z}a+\mathcal{K}_{z}a})+h.c.)\Big]_{min}\Delta y^{2}+\dots,
\end{equation}
where we expanded around the SUSY minimum $\{u=u_{\star}, z=z_{\star}\}$ and $\Delta y\equiv y-y_{\star}$. The ellipses stand for terms due to backreaction of $z$ , which will be studied in detail in Sec.~\ref{sec:branalytical}. It is now apparent that flatness of the potential cannot be ensured by tuning $a$ alone. Instead, we also require $|\pd_z a|$ to be sufficiently small. It is important to notice that small $|a|$ does not imply small $|\pd_z a|$. In the context of string theory compactifications with flux, parameters can be made small by tuning: various terms which are not small individually contribute to $a(z)$ and can be made to cancel up to a small remainder. However, this cancellation will generically not occur in $\pd_z a$. Requiring a small value for $|\pd_z a|$ hence introduces an additional tuning. The analysis can be easily generalised to the case of more than two moduli. For every additional modulus $z^j$ we also require $|\pd_{z^j}a|$ to be sufficiently small. Therefore, for $n$ moduli $z^i$ we have to tune $(n+1)$ quantities.

It is easy to see that one cannot get away with fewer tunings. The argument is as follows. Find the basis in which the K\" ahler metric is diagonal. In this basis the inflationary potential is a sum of positive terms (in essence, the mixed terms $\sim \mathcal{K}^{z\bar{z}},\mathcal{K}^{z\bar{u}}$ in \eqref{eq:fpotential1} disappear). Therefore, in order to achieve a flat direction, each contribution has to be tuned small. One then has to tune $(n+1)$ different combinations of $a$ and $\partial_{z^i}a$, $i=1,\dots,n$.  These combinations involve elements of the original inverse K\"ahler metric as coefficients. It is conceivable that these terms could take small values in some region of the moduli space. This corresponds to special geometries of moduli space where, at particular points, certain elements of the metric blow up. Since we do not know whether such situations can occur, in particular given that one complex structure modulus (the inflaton) must be stabilised in the large complex structure limit, we choose not to consider this option in the following. Thus, for the case of $n$ moduli $z^{i}$, we require $|a|$, $|\pd_{z^1} a|$, $|\pd_{z^2} a|$, \ldots $|\pd_{z^n} a|$ to be small. 

It follows that models of $F$-term monodromy inflation are more severely tuned than initially anticipated. One aim of this paper is to estimate the number of string vacua with the desirable properties for $F$-term axion monodromy inflation. While our estimate will be fairly rough, it will be sufficient to decide whether there is still a landscape of acceptable vacua. We will address this issue in section \ref{sec:landscape}.

\begin{figure}
	\centering
		\includegraphics[width=0.50\textwidth]{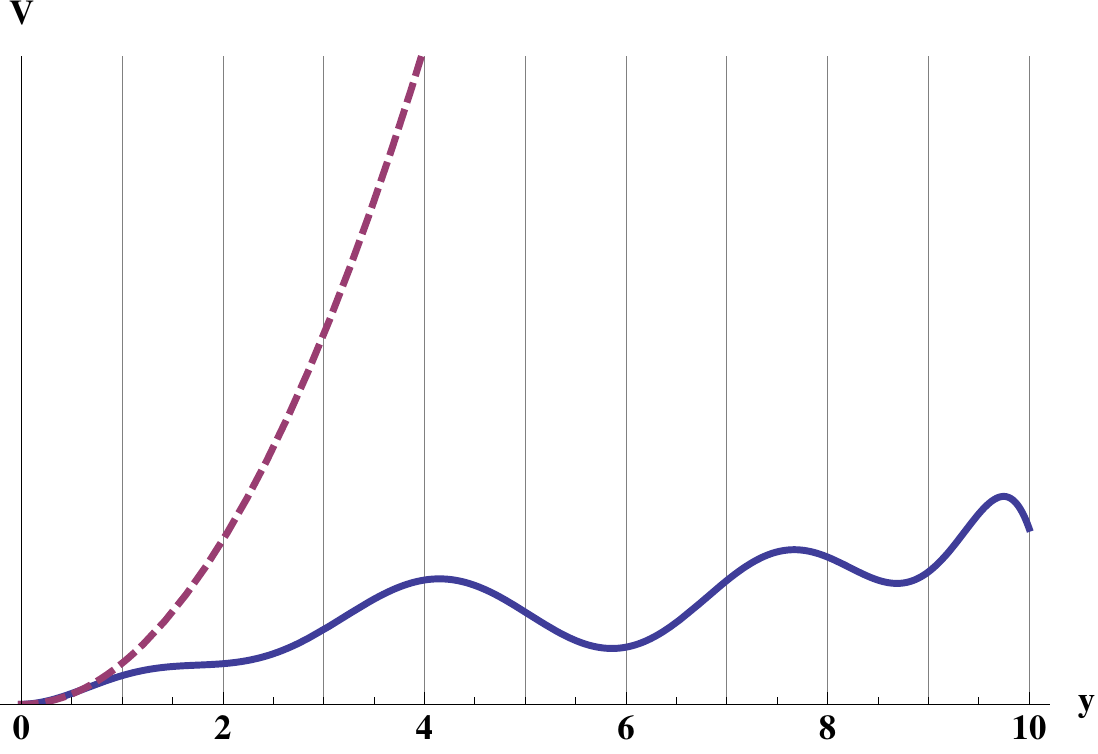}
	\caption{`Naive' inflaton potential (dashed red line) and a possible effective inflaton potential after backreaction is taken into account (solid blue line). Note that the effective inflaton potential is not automatically sufficiently flat over transplanckian regions to realise large field inflation.}
\label{fig:backreact01}
\end{figure}

There is also a second problem which we address in this paper. In the above analysis we saw the importance of keeping the complex structure moduli $z$ dynamical. In this setting we can then also address the question of backreaction of the potential on the moduli $z$ \cite{10114521, 14053652}. In particular, notice that the term $a(z) u \subset W$, while giving rise to the inflaton potential, also corresponds to a cross-coupling between $z$ and $u$. The danger then is that for large field displacements of $u$, as required in models of large field inflation, the moduli $z$ could be significantly displaced from their values at the global minimum. The consequences are as follows. While the potential \eqref{eq:fpotential1} is rising monotonically in the $y$-direction, this behaviour could change dramatically once the moduli $z$ are allowed to adjust. In particular, it is not clear that the flattest direction away from the global minimum will rise monotonically over a transplanckian field space. Instead, after an initial rise one might encounter a series of local minima. This is illustrated in figure \ref{fig:backreact01}. 

Consequently, we will have to examine backreaction of complex structure moduli $z$, which we will do both analytically and numerically. In particular, we will determine whether the tuning of the parameters above is sufficient to control backreaction. We will also study the resulting effective inflaton potential and check whether it is suitable to realise inflation. This is the subject of sections \ref{sec:branalytical} -- \ref{sec:tuneless} and \ref{sec:brnum}.

However, before we embark on these analyses we review how models of type \eqref{eq:themodel} can arise in string compactifications. In addition, we check whether the required tunings can be implemented in type IIB orientifolds and F-theory 4-folds.

\subsection{A No-Go-Theorem for type IIB orientifolds at weak coupling}
\label{sec:nogo3}
As argued in the previous subsection, any successful large field inflation model based on \eqref{eq:themodel} with a complex structure modulus $u$ in the large complex structure (LCS) regime requires a flux tuning of not only $|a|$ but also of all $|\partial_{z^ i}a|$ and $|\partial_{S}a|$ at the minimum with $S = i/g_s +C_0$ being the axio-dilaton.\footnote{In the orientifold case $S$ enters the $F$-term scalar potential similarly to the complex structure moduli. Thus also $|\partial_{S}a|$ has to be tuned to a small value.} Let $X$ be the orientifold on which we wish to realise large field inflation. We denote by $z^I$, $I=0,\dots,n$ the $h^{2,1}_-(X)=n+1$ complex structure moduli with the inflaton being $z^0 \equiv iu$. Throughout the whole paper upper-case indices run from $0$ to $n$, while lower-case indices run from $1$ to $n$. In the orientifold case, the most general form of the superpotential $W$ with $u$ in the LCS limit is given by 
\begin{equation}\label{superpotential threefold}
W=w(S,z)+a(S,z)u + \frac{1}{2}b(S,z)u^2 + \frac{1}{3!}c(S)u^3.
\end{equation}
Here, $z$ denotes all the $z^i$, $i=1,\dots,n$.  We now briefly show that $a$ and $b$ depend on $S$ and the $z^i$, while only $S$ enters $c$. One starts from the Gukov-Vafa-Witten potential \cite{0105097}
\begin{equation}
W = \int_X (F_3-SH_3)\wedge \Omega_3 \ ,
\end{equation}
where $F_3$ and $H_3$ are the type IIB three-form fluxes and $\Omega_3$ is the holomorphic $(3,0)$-form on the threefold $X$. After flux quantisation one can write 
\begin{equation}\label{GKV threefold}
W=(N_{F}-SN_{H})^{\alpha}\Pi_{\alpha}
\end{equation}
with the flux vectors $N_F, N_H$ and the period vector $\Pi$, which is given by \cite{14050283, 9403096}
\begin{equation}\label{period vector threefold}
\Pi_{\alpha} = \begin{pmatrix}
 1 \\
 z^{I} \\
\frac{1}{2}\kappa_{IJK}z^{J}z^{K} + f_{IJ}z^J + f_I  + \sum_{p}A_{I p} e^{-\sum_{J}b_{pJ}z^{J}} \\
-\frac{1}{3!}\kappa_{IJK}z^{I}z^{J}z^{K} + f_Iz^I + g + \sum_p B_p e^{-\sum_{J}\tilde{b}_{pJ}z^{J}}
\end{pmatrix} \ .
\end{equation}
Here, $\kappa_{IJK}$ ($I,J,K =0,\dots,n$) denote the triple intersection numbers of the 4-cycles of the mirror dual CY threefold $\tilde{X}$. Moreover, the flux index $\alpha$ runs from $\alpha = 1,\dots, 2h^{2,1}_-(X)+2=2n+4$ in our case. By stabilising $u$ in the LCS limit, i.e.~$\mathrm{Re}(u) \gtrsim \mathcal{O}(1)$, the (worldsheet) instanton terms $e^{-2\pi u}$ are suppressed. Instanton terms containing $z^i$ but not $u$ are not suppressed, but they only enter $w(S,z)$. Not much is known about the subleading terms $f_{IJ}, f_I $ and $g$. In examples we are aware of, those terms turn out to be zero or half-integers (see e.g.~\cite{9403096, 9406055, 10105792, 14050291}) and hence, as we will explain, they will be irrelevant for the arguments below. Therefore we drop those terms in the following. Then, from \eqref{GKV threefold} it follows that $u$ enters $W$ up to power three, as stated in \eqref{superpotential threefold}. Clearly, $S$ only appears linearly in $W$. In particular, $c$ cannot depend on the $z^i$ because only $\kappa_{000}u^3$ can contribute to $c$, and thus 
\begin{equation} \label{c(S)}
c(S) \sim (m+nS)
\end{equation}
with $m, n \in \mathbb{Z}$. Similarly, from \eqref{GKV threefold} together with \eqref{period vector threefold} one can easily see that $a$ and $b$ depend on the $z^i$ and $S$ as follows:
\begin{equation} \label{a(S,z)}
a(S,z) \sim (\alpha + \beta S + \gamma_{i}z^ i + \lambda_{i}Sz^i + \zeta_{ij}z^iz^j + \xi_{ij}Sz^iz^j )
\end{equation} 
and
\begin{equation} \label{b(S,z)}
b(S,z) \sim (\tilde{\alpha} + \tilde{\beta} S + \tilde{\gamma}_{i}z^ i + \tilde{\lambda}_{i}Sz^i)
\end{equation} 
with integers $\alpha, \beta, \gamma_{i}, \lambda_i, \zeta_{ij}, \xi_{ij}, \tilde{\alpha}, \tilde{\beta}, \tilde{\gamma}_i, \tilde{\lambda}_i$.
\\
Note that for successful inflation we not only have to tune $|a|$ and its derivatives small (as explained in the previous section), but also $|b|, |\partial_{z^ i}b|, |\partial_{S}b|, |c|, |\partial_{S}c|$ have to be small quantities at the minimum. This can be seen as follows. First of all, as we only want to break the shift symmetry in $u$ weakly $|a|$, $|b|$ and $|c|$ need to be small. However, the scalar potential will receive further contributions which break the shift symmetry. In particular, we have $D_iW \supset ((\partial_{z^i} a)u +(\partial_{z^i} b) u^2/2)$ and $D_SW \supset ((\partial_S a) u +(\partial_S b) u^2/2 + (\partial_S c) u^3/3!)$, which do not obey the shift symmetry. Thus, the derivatives $|\partial_{z^ i}a|, |\partial_{S}a|, |\partial_{z^ i}b|, |\partial_{S}b|, |\partial_{S}c|$ indeed need to be tuned small as well. We will show in the following that these tunings either make it impossible to stabilise $g_s$ in the perturbative regime or to stabilise $\mathrm{Re}(u)$ successfully. 
\\
For this, we first prove the following statement:  

\bigskip
\noindent Statement 1: \textit{In the perturbative regime one cannot tune $|c(S)|$ small, i.e.~$|c(S)|<\epsilon$ with $\epsilon \ll 1$, as long as $c(S) \neq 0$.}
\bigskip
\\ 
By \eqref{c(S)}, the tuning condition $|c(S)|<\epsilon$ translates into 
\begin{equation}
|m+nS|<\epsilon,
\end{equation}
 where $m,n \in \mathbb{Z}$. Therefore, both the real and the imaginary part of $m+nS$ have to be as small as $\epsilon$ individually. Thus, $|n \ \mathrm{Im}(S)|<\epsilon$ for the imaginary part. However, because of $S=i/g_s +C_0$, it follows that
\begin{equation}
|n \ \mathrm{Im}(S)| = \frac{|n|}{g_s} < \epsilon.
\end{equation} 
If $n \neq 0$, then $g_s > |n|/ \epsilon \gg 1$. Hence, in this case it is impossible to stabilise $g_s$ in the perturbative regime. Since $n \in \mathbb{Z}$, one cannot simply tune $n$ small. Therefore, one can only evade $g_s \gg 1$ if we choose $n=0$. But then, $|c(S)|=|m|<\epsilon$, i.e.~$m=0$. This implies that $c(S)$ has to vanish identically.  
\\
This observation allows to go even one step further and to state and prove the following:

\bigskip
\noindent Statement 2: \textit{On any CY threefold with $\kappa_{000}\neq 0$ or $\kappa_{i00} \neq 0$ for some $z^{i}$, the tuning requirements for large field inflation imply that the string coupling is stabilised at $g_s \gg 1$.}
\bigskip
\\
The proof is as follows. We have to tune all the parameters $a,b,c$ and their derivatives as small as $\epsilon \ll 1$. Statement 1 shows that being in the perturbative regime requires $c\equiv 0$. 
\\
There are two possibilities to make $c$ vanish identically. One could choose a CY threefold with $\kappa_{000}=0$ or turn off the last entries of the flux vectors (i.e.~choosing the flux numbers $(N_F)_{2n+4}$ and $(N_H)_{2n+4}$ to zero). Let us first consider the latter possibility. From \eqref{period vector threefold} one can see that turning off these flux numbers indeed prevents $\kappa_{000}u^3$ from entering $W$.  However, one would also simultaneously forbid the terms $\sim \kappa_{i00} z^iu^2$, i.e.~$\tilde{\gamma}_i=\tilde{\lambda}_i=0$ for all $i$. Thus, $b(S,z)=b(S)$ (see \eqref{b(S,z)}), i.e.~it then has the same moduli-dependence as $c$. In analogy, by using statement 1, we can then infer that $b \equiv 0$. Furthermore, $(N_F)_{2n+4}=(N_H)_{2n+4}=0$ implies that $\zeta_{ij}=\xi_{ij}=0$ for all $i,j$. We see that \eqref{a(S,z)} becomes 
\begin{equation}
a(S,z) \sim (\alpha + \beta S + \gamma_{i}z^ i + \lambda_{i}Sz^i),
\end{equation} 
and therefore $\partial_ja(S,z) \sim (\gamma_j + \lambda_j S)$, $\gamma_j, \lambda_j \in \mathbb{Z}$ for all $j$.  Consequently, the tuning condition $|\partial_ja(S,z)|<\epsilon$ translates into $|\gamma_j + \lambda_j S|<\epsilon$ and again, by statement 1 we are forced to choose $\gamma_j = \lambda_j=0$, and we are left with $a(S)\sim (\alpha + \beta S)$. Once more, $|a(S)|<\epsilon$ yields $a=0$ by statement 1.\footnote{If $f_{IJ}$ is an integer or a half-integer it does not influence the argument. However, if there are cases in which $f_{IJ}$ can be irrational or a sufficiently complicated fraction, there is a chance to evade the conclusion $a(S)=0$. Instead one could use $f_{IJ}$ to tune the whole expression $a(S)$ small. Since we are not aware of examples in which the terms $f_{IJ}$ are irrational numbers or complicated fractions, we do not consider this possibility any further.}  

However, even for $\kappa_{000}=0$ we are forced to set $a=b \equiv 0$ in order to avoid $g_s \gg 1$. The requirement $|\partial_{k}b|<\epsilon$ yields $|\tilde{\gamma}_k + \tilde{\lambda}_kS|<\epsilon$ with $\tilde{\gamma}_k, \tilde{\lambda}_k \in \mathbb{Z}$. By statement 1 one must have $\tilde{\gamma}_j=\tilde{\lambda}_j=0$ for all $j$. Then, again, the condition $|b|<\epsilon$ forces us to choose $\tilde{\alpha}=\tilde{\beta}=0$ due to statement 1. Hence, $b$ has to vanish identically, too. Since, by assumption, $\kappa_{i00} \neq 0$ for some $z^{i}$ if $\kappa_{000}=0$, one cannot avoid choosing $(N_F)_{2n+4}$ and $(N_H)_{2n+4}$ to be zero, otherwise $b \neq 0$. This then implies  $\zeta_{ij}=\xi_{ij}=0$.  By repeating the above arguments, we find $a \equiv 0$ or $g_s > 1/\epsilon$. This proves statement 2. 

Obviously, if we consider a CY threefold with $\kappa_{000}=0=\kappa_{i00}$ for all $z^i$ (K3-fibrations admit such triple intersection numbers), then we have $b=c\equiv 0$, but generically $\zeta_{ij},\xi_{ij} \neq 0$, because no fluxes have to vanish. In this case, it seems possible to stabilise $g_s$ in the perturbative regime. However, it is then not clear how to stabilise $\mathrm{Re}(u)$ successfully. Note that a CY threefold with the above triple intersection numbers yields a K\"ahler potential of the form 
\begin{equation}
\mathcal{K}_{cs} = - \ln \left(A(z) +B(z)(u+\bar{u}) \right)
\end{equation} 
with $A, B$ being functions of the remaining complex structure moduli. Then, the contribution 
\begin{equation}
e^{\mathcal{K}_{cs}}V_{\mathrm{LVS}} \sim - e^{\mathcal{K}_{cs}} \frac{|W|^2}{\mathcal{V}^3}
\end{equation}
from the LVS-potential, which dominates the $F$-term potential for $\mathrm{Re}(u)$, does not admit a minimum for $\mathrm{Re}(u)$ in the regime where $A+B(u+\bar{u})>0$, but rather shows a runaway behaviour. This issue is rooted in the simple structure of the K\"ahler potential. Note that an analogous problem occurs in inflation models with the universal axion, where the string coupling $g_s$ needs to be stabilised. Consequently, large field inflation with a complex structure modulus in the LCS limit cannot be realised on CY threefolds with $\kappa_{000}=0=\kappa_{i00}$ for all $z^i$. Together with statement 2, we summarise our findings as follows (and refer to it as the no-go theorem henceforth):
 
\bigskip
\noindent \textit{For any orientifold with at least one complex structure modulus $u$ in the large complex structure limit, at least one of the following three conditions cannot be satisfied: \begin{enumerate} \item The coefficients in front of the inflaton field $u$ in the superpotential $W$ and their derivatives are tuned sufficiently small to allow for inflation. 
\item The string coupling $g_s$ is stabilised in the perturbative regime. 
\item $\mathrm{Re}(u)$ can be stabilised using the classical supergravity $F$-term scalar potential.  
\end{enumerate}}
\bigskip
Note that possible scenarios where only condition 3 is violated deserve more detailed investigation in future work. For instance, certain uplifting scenarios or a mild interference with K\"ahler moduli stabilisation could turn out to be a loophole concerning the problems in stabilising $\mathrm{Re}(u)$ that were outlined above.  

This no-go theorem can be evaded by considering Calabi-Yau fourfolds as the starting point for the subsequent analysis.

\subsection{Calabi-Yau fourfolds in a partial Large Complex Structure regime}
\label{sec:4foldsetup}
As explained, the no-go theorem forces us to work with Calabi-Yau \textit{fourfolds} $X$, whose complex structure moduli are denoted by $u \equiv z^0$ and $z^i$, $i = 1,\dots,n$, where $n = h^{3,1}(X)-1$. Useful references for this section are \cite{0105097, 08031194}. Again, $u$ labels the complex structure modulus in the large complex structure regime which contains the inflaton field.

The superpotential $W$ can be computed directly from the Gukov-Vafa-Witten potential \cite{0105097}
\begin{equation}
W = \int_{X}G_4 \wedge \Omega_4 \ ,
\end{equation}
with $G_4$ and $\Omega_4$ being the 4-form flux and the holomorphic 4-form on $X$, respectively. After flux quantisation this gives
\begin{equation}
W =  N^{\alpha} \Pi_{\alpha}\ ,
\end{equation}
where $N$ is flux vector and $\Pi$ denotes the period vector with $\alpha = 1,\dots,b_4(X)$. Schematically, $\Pi$ has the following structure \cite{14050283, 13023760, 9403096}: 
\begin{equation}
\Pi_{\alpha} \sim \begin{pmatrix}
 1 \\
 z^I \\
\kappa_{IJKL} z^Kz^L + \mathrm{Inst}(u,z) \\
\kappa_{IJKL}z^Jz^Kz^L  + \mathrm{Inst}(u,z)  \\
\kappa_{IJKL} z^Iz^Jz^Kz^L + \mathrm{Inst}(u,z) 
\end{pmatrix} \ ,
\end{equation}
where $\kappa_{IJKL}$ denote the intersection numbers of the 6-cycles of the mirror dual CY fourfold, and $\mathrm{Inst}(u,z)$ summarises various instanton terms, depending on $u$ and all the $z^i$. 

In general, $W$ is a holomorphic function in $u$ and the remaining complex structure moduli. In this work we wish to only consider superpotentials where $u$ appears at most linearly: $W=w + a u$. The main motivation behind this restriction is to keep the analyses in the following chapters simple. In principle, the study of backreaction performed in this work should also be possible for models with a more complicated superpotential, but we leave this for future studies.

In the following we will argue how a superpotential linear in $u$ can be obtained. One obstruction to this is the presence of non-perturbative terms of the form $\sim e^{-2\pi u}$ in $\Pi$. As before, by working in the LCS regime where $u$ is large we can ensure that all non-perturbative terms containing $u$ are exponentially suppressed. Note that we do not require that all moduli $z^I$ need to be in the LCS regime: we only require a subset including $u$ to be at LCS, which we refer to as `partial large complex structure'. Then, at this stage, $u$ can arise at most as $u^4$ in $W$. 

In order to achieve a superpotential of the form $W=w(z)+a(z)u$, we assume $X$ to have intersection numbers $\kappa_{0000}=0=\kappa_{i000}$ for all $z^i$. Hence, cubic or quartic terms in $u$ are prohibited by the geometry of $X$. All terms which potentially give rise to quadratic terms in $u$ need to be set to zero by a corresponding flux choice.  For instance, the last component of $\Pi$ contains $\kappa_{ij00}z^iz^ju^2$, which does not necessarily vanish, and thus the last component of $N$ must be chosen to be zero. Since the Betti number $b_4(X)$ does not only receive contributions from $h^{3,1}(X)$ but also from $h^{2,2}(X)$, we expect that the available number of flux parameters exceeds the number of required tunings. For instance, if $X$ is an elliptic fibration over $\mathbb{C}P^3$ one obtains $h^{3,1}(X)=3878$, $h^{2,2}(X)=15564$ and hence $b_4(X)=23320$ \cite{08031194}. Thus, in this example one has many more flux numbers than complex structure moduli.      

We now want to write down the tuning conditions explicitly and argue that these requirements can be satisfied in principle. Using the notation $\vec{z} \equiv (z^1,\dots,z^{n})$, we can write $a(\vec{z})$ schematically as 
\begin{equation}
a(\vec{z}) \sim (m+\vec{n}^{t}\vec{z}+\vec{z}^{t}\textbf{N}\vec{z})
\end{equation}
with $m \in \mathbb{Z}$, $\vec{n} \in \mathbb{Z}^{n}$ and $\textbf{N}$ being an integer valued matrix. The tuning condition on the derivatives of $a(\vec{z})$ is $|\nabla a| \lesssim \epsilon \simeq 0$. This gives two real equations 
\begin{align} \label{tuning condition 1}
2\textbf{N}\vec{v} & \simeq - \vec{n}, \\ 
\textbf{N}\vec{w} & \simeq 0,
\end{align}
where $\vec{v} = \mathrm{Re}\vec{z}$ and $\vec{w} = \mathrm{Im}\vec{z}$. Inserting these results into $a(\vec{z})$, we find 
\begin{equation}
a(\vec{z}) \sim (m + \frac{1}{2}\vec{n}^t\vec{v} + \frac{1}{2}i\vec{n}^t\vec{w}) \ .
\end{equation}
As we need to tune $|a(\vec{z})| \lesssim \epsilon \simeq 0$ we also require
\begin{align}
|\vec{n}^t\vec{w}| & \simeq 0 \\
\left|m + \frac{1}{2}\vec{n}^t\vec{v}\right| & \simeq 0. \label{tuning condition 4}
\end{align}
A solution to these four conditions is as follows. Suppose $\det \textbf{N} \neq 0$, then $\vec{w}\simeq 0$, i.e.~the second and the third conditions are satisfied. The first condition \eqref{tuning condition 1} can be solved for $\vec{v}$ and plugged into \eqref{tuning condition 4} to get the requirement 
\begin{equation}
m \simeq \frac{1}{4} \vec{n}^t (\textbf{N}^t)^{-1}\vec{n}.
\end{equation} 
This can be satisfied easily if e.g.~$\det \textbf{N} = \pm 1$, since in this case $\textbf{N}^{-1}$ is again integer valued. Of course, there can also be solutions to the tuning conditions for $\det \textbf{N} = 0$, but we do not study them any further since our intention was to show that one can in principle satisfy the tuning  requirements. 

We now turn to the K\"ahler potential $\mathcal{K}_{\mathrm{cs}}$ for the complex structure moduli. It can be determined from the period vector $\Pi$ as 
\begin{equation}
\mathcal{K}_{cs} = - \ln \left(\Pi_{\alpha}(z,u)Q^{\alpha\bar{\beta}} \overline{\Pi_{\beta}}(\bar{z},\bar{u}) \right)
\end{equation} 
with the intersection matrix $Q_{\alpha\bar{\beta}}$. Most importantly, since $u$ is taken to be in the LCS regime, it appears only as $u+\bar{u}$ in the K\"ahler potential. Consequently, the K\"ahler potential for the complex structure moduli is indeed of the form $\mathcal{K}_{cs} = \mathcal{K}_{cs}(z,\bar{z},u+\bar{u})$, as stated in \eqref{eq:themodel}. From the structure of the period vector it is also evident that $\mathcal{K}_{cs}$ can in principle contain a polynomial in $(u+\bar{u})$ of degree four (at most). Since, for simplicity, we consider $\kappa_{0000}=0=\kappa_{i000}$ for all $z^i$, we have in fact a quadratic polynomial in $(u+\bar{u})$ in the logarithm of the K\"ahler potential. However, note that we do not rely on the specific structure of $\mathcal{K}_{cs}$ for the subsequent analysis. The crucial point is the existence of the shift-symmetry of $\mathcal{K}_{cs}$ (under $u \rightarrow u + i \alpha$), which is a necessary requirement to evade the $\eta$-problem.\footnote{In addition, we require the existence of a point where $\pd_u \mathcal{K} =0$. This will allow us to stabilise $\textrm{Re}(u)$ through $\mathcal{K}$ and $w$ only (see \cite{14043711} for more detail). We can then ensure that $\textrm{Re}(u)$ is parametrically heavier than the inflaton $\textrm{Im}(u)$, which only acquires a mass through $au \subset W$. A K\"ahler potential with a quadratic polynomial in $(u+\bar{u})$ inside the logarithm is sufficient to stabilise $\mathrm{Re}(u)$ through $\mathcal{K}$ and $w$ only.} Again, to arrive at a K\"ahler potential with one shift-symmetric direction, we do not require all complex structure moduli to be at LCS: only a subset of complex structure moduli containing $u$ has to be large. As before, `partial large complex structure' is sufficient. Overall, this leaves F-theory 4-folds as a promising starting point for models of $F$-term axion monodromy inflation.  

For the sake of simplifying the notation, we henceforth abbreviate $f_I \equiv \partial_{z^I}f$, $I=0,\dots,n$ and $f_i \equiv \partial_{z^i}f$, $i=1,\dots,n$ for any function $f$.


\subsection{Backreaction and the effective inflaton potential}
\label{sec:branalytical}
In this section we will study the backreaction on the complex structure moduli $z^i$, $\bar{z}^i$ as well as on $x\equiv \text{Re}(u)$, if we displace $y\equiv \text{Im}(u)$ by some finite distance $\Delta y$ from the minimum. In particular, we will derive the effective inflaton potential once backreaction is taken into account. 

The starting point are a K\"ahler potential and a superpotential of the form
\begin{equation}
\label{eq:superkaehler}
W=w(z)+a(z)u, \quad \mathcal{K}\equiv \mathcal{K}(z, \bar{z}, u+\bar{u}) \ ,
\end{equation}
from which we can determine the $F$-term scalar potential
\be
\label{eq:SUGRAV02} V = e^{\mathcal{K}} (\mathcal{K}^{I \bar{J}} D_I W \overline{D_J W}) \ .
\ee
Most importantly, we do not assume that any of the $z^i$ are integrated out. On the contrary, we take all $z^i$ as well as $u$ to be dynamical. To quantify backreaction the strategy is as follows. We expand the potential in $\delta z^i$, $\delta \bar{z}^i$ and $\delta x$ to quadratic order about the minimum. As long as the displacements $\delta z^i$, $\delta \bar{z}^i$ and $\delta x$ remain small during inflation this expansion is a good approximation to the full potential and higher order terms can be ignored. For every value of $\Delta y$ the potential is then a quadratic form in the displacements of the remaining fields. As such, it admits a global minimum at each value of $\Delta y$ for some $\delta z^i (\Delta y)$, $\delta \bar{z}^i (\Delta y)$ and $\delta x (\Delta y)$, which we calculate explicitly. In the following we will show that the displacements $\delta z^i (\Delta y)$, $\delta \bar{z}^i (\Delta y)$ and $\delta x (\Delta y)$ are indeed small for a wide range in $\Delta y$ such that our analysis is self-consistent. By substituting these solutions into the expression for the potential we can then derive the effective inflaton potential.

We now perform the steps outlined above explicitly. To begin, we wish to expand the scalar potential \eqref{eq:SUGRAV02} to quadratic order in $x$, $z^j$ and $\zb^j$ about their values at the minimum. For this, it will be sufficient to expand the covariant derivatives $D_I W$ to first order. Indeed the inverse K\"ahler metric and the exponential prefactor do not contribute at quadratic order, as shown by varying the $F$-term potential twice:
\begin{align}
\nonumber \delta^{2} V_{F}&=\delta^{2}\Big(e^{\mathcal{K}}\mathcal{K}^{I\bar{J}}\Big)\Big[D_{I}W\overline{D_{J}W}\Big]_{min}+\delta\Big(e^{\mathcal{K}}\mathcal{K}^{I\bar{J}}\Big)\delta\Big(D_{I}W\Big)\Big[\overline{D_{J}W}\Big]_{min}\\
&+\delta\Big(e^{\mathcal{K}}\mathcal{K}^{I\bar{J}}\Big)\delta\Big(\overline{D_{J}W}\Big)\Big[D_{I}W\Big]_{min}+\Big[e^{\mathcal{K}}\mathcal{K}^{I\bar{J}}\Big]_{min}\delta^{2}\Big(D_{I}W\overline{D_{J}W}\Big),
\end{align}
and imposing the minimum condition $D_{I} W=0$.

The covariant derivatives are given by:
\begin{align}
\label{eq:covariant}
\nonumber D_u W&= a+\mathcal{K}_u(w+ax+iay),\\
D_{z^i} W&= w_i+a_{i}(x+iy)+\mathcal{K}_{i}(w+ax+iay).
\end{align}
Recall that a subscript $i$ corresponds to a derivative w.r.t.~$z^i$: $f_i \equiv \pd_{z^i} f$. The values $u_{\star}, z_{\star}$ of the complex structure moduli at the minimum are found by imposing the conditions:
\begin{equation}
D_{u}W=0, \quad D_{z^{i}} W=0.
\end{equation}
The latter can be solved in terms of the derivatives of the K\" ahler potential at the minimum:
\begin{align}
\label{eq:kconstraint}
\nonumber D_{u} W&=0 \Rightarrow \mathcal{K}_{u}|_{\star}=-\frac{a}{w+au}\Big{|}_{\star}\\
D_{z^{i}}W&=0 \Rightarrow \mathcal{K}_{i}|_{\star}=-\frac{w_i+a_{i}u}{w+au}\Big{|}_{\star} \quad .
\end{align}
We now write $z^j=z_{\star}^j + \delta z^j$ and $u= u_{\star} + \delta x + i \Delta y$ and expand \eqref{eq:covariant} to linear order in $\delta x$, $\delta z^j$ and $\delta \zb^j$. Note that we will not perform an expansion in $\Delta y$. On the contrary, our result will be exact in $\Delta y$. This is absolutely crucial as $\Delta y$ will take transplanckian values during inflation and is not a small quantity. In the following it will also be useful to absorb the term $au_{\star}$ into a quantity $w_{*}$:
\be
w_{*} \equiv W(z, u_{\star}) = w(z) + a (z) u_{\star} \ .
\ee
Expanding \eqref{eq:covariant} to linear order in $\delta x$, $\delta z^j$ and $\delta \zb^j$ we find:
\begin{align}
\label{eq:linearcovu}
\nonumber D_{u} W & =\Big[a_{j}+\mathcal{K}_{uj}w_{*}+\mathcal{K}_{u}w_{*j}+i(\mathcal{K}_{uj}a+\mathcal{K}_{u}a_{j})\Delta y\Big]_{\star}\delta z^j\\
\nonumber &+ \Big[\mathcal{K}_{u\bar{j}}w_{*}+ia\mathcal{K}_{u\bar{j}}\Delta y\Big]_{\star}\delta\bar{z}^j\\
&+ \Big[\mathcal{K}_{u}a+\mathcal{K}_{ux}w_{*}+i \mathcal{K}_{ux}a\Delta y\Big]_{\star}\delta x+ i\Big[\mathcal{K}_{u}a\Big]_{\star}\Delta y + O(\delta^{2}),\\
\label{eq:linearcovz}
\nonumber D_{z^i} W & = \Big[w_{*ij}+\mathcal{K}_{ij}w_{*}+\mathcal{K}_{i}w_{*j}+i(a_{ij}+\mathcal{K}_{ij}a+\mathcal{K}_{i}a_{j})\Delta y\Big]_{\star}\delta z^j\\
\nonumber &+\Big[\mathcal{K}_{i\bar{j}}w_{*}+i\mathcal{K}_{i\bar{j}}a\Delta y\Big]_{\star}\delta\bar{z}^j\\
&+\Big[ a_i + \mathcal{K}_{ix}w_{*}+\mathcal{K}_{i}a+i\mathcal{K}_{ix}a\Delta y\Big]_{\star}\delta x + i\Big[a_{i}+\mathcal{K}_{i}a\Big]_{\star}\Delta y+O(\delta^{2}).
\end{align}
Here we used the subscript $\star$ to make it explicit that the quantities in square brackets are evaluated at the minimum, but we will suppress it in what follows.

If the displacements $\delta z, \delta\bar{z}, \delta x $ are small, the leading term in the potential is quadratic in $\Delta y$. This term is therefore the \emph{naive} inflationary potential and reads:
\begin{equation}
\label{eq:naive}
V_{naive}\sim \left[ \mathcal{K}^{u\bar{u}}|\mathcal{K}_{u}a|^{2}+\mathcal{K}^{i\bar{j}} (a_{i}+\mathcal{K}_{i}a) \overline{(a_{j}+\mathcal{K}_{j}a)}+(\mathcal{K}^{u\bar{j}}(\mathcal{K}_{u}a)(\overline{a_{j}+\mathcal{K}_{j}a})+h.c.) \right](\Delta y)^{2}
\end{equation}
In order for $\Delta y$ to be a suitable direction for inflation, we require that the naive potential is almost flat. From (\ref{eq:naive}), this requirement is satisfied if $|\mathcal{K}_{u}a|$ and $|a_{j}+\mathcal{K}_{j}a|$ are small. This can be achieved by tuning all the parameters $|a|, |a_{j}| \ll 1$.\footnote{In the case of $k$ complex structure moduli entering $a$, these are $k+1$ tunings. As we have discussed in Sec. (\ref{sec:theproblem}), one cannot get away with fewer tunings.} In order to obtain compact expressions, we introduce the following quantities:
\begin{align}
\label{eq:small}
\nonumber \eta_u&=i \mathcal{K}_u a\\
\eta_j&= i (a_{j}+\mathcal{K}_{j}a).
\end{align}
At this point, it is important to notice that \eqref{eq:kconstraint} imposes $\mathcal{K}_{u}\sim a$. The latter implies that $\mathcal{K}_{u}$ is as small as $a$ at the minimum, while $\mathcal{K}_{i}$ and the elements of the K\" ahler metric are not parametrically small. Introducing the small parameter
\be
\label{def:epsilon}
\epsilon \equiv |a| \ ,
\ee 
it follows that $\eta_{u}\sim \epsilon^{2}$ while the second term in $\eta_{j}$ is only proportional to $\epsilon$. In this and the following subsection we assume that $a_{i}$ is tuned in such a way that $\eta_{i}\sim \epsilon^{2}$ as well. Under these assumptions $\eta_{u}$ and $\eta_{i}$ are parametrically of the same size. This turns out to be useful for our explicit computations. We discuss the generic case of hierarchical $\eta$'s in section \ref{sec:tuneless}.

We can now simplify our expressions \eqref{eq:linearcovu}. We will later show that the displacements $\delta x, \delta z^j, \delta \zb^j$ are small to the extent that $\eta_u, \eta_j$ are small. In particular, when $\eta_u \sim \eta_j \sim \epsilon^2$ we will find that $\delta x \sim \delta z^j \sim \delta \zb^j \sim \epsilon^2$. It follows that e.g.~$a_j \delta z ^j \sim \epsilon^{3}$ while $\mathcal{K}_{uj}w_{*} \delta z^j \sim \epsilon^2$. To simplify further, we can then neglect those terms in \eqref{eq:linearcovu} that are smaller than $O(\epsilon^2)$. Let us be more precise about the latter statement. In \eqref{eq:linearcovu} and \eqref{eq:linearcovz} there are terms of the form $O(\epsilon^{3})\Delta y$. Those terms are negligible compared to those of $O(\epsilon)$ as long as $\Delta y\ll\epsilon^{-1}$. We shall therefore restrict the field displacement to $0<\Delta y\ll\epsilon^{-1}$. This is also motivated by the following argument. In order not to interfere with K\" ahler moduli stabilisation we need to impose $au\sim {\epsilon}u\ll w$ in \eqref{eq:superkaehler}. This constraint then implies the same restriction on the field range. We thus arrive at:
\begin{align}
\label{eq:linearepcovu}
 D_{u} W & \simeq \Big[\mathcal{K}_{uj}w_{*}\Big]\delta z^j+\Big[\mathcal{K}_{u\bar{j}}w_{*}\Big]\delta\bar{z}^j+ \Big[\mathcal{K}_{ux}w_{*}\Big]\delta x+ \eta_{u}\Delta y,\\
\label{eq:linearepcovz}
D_{z^i} W & \simeq \Big[w_{*ij}+\mathcal{K}_{ij}w_{*}+\mathcal{K}_{i}w_{*j}+i(a_{ij})\Delta y\Big]\delta z^j+\Big[\mathcal{K}_{i\bar{j}}w_{*}\Big]\delta\bar{z}^j \\
\nonumber &+\Big[\mathcal{K}_{ix}w_{*}\Big]\delta x + \eta_{i}\Delta y \quad .
\end{align}
Note that at leading order $D_u W \sim D_{z^i} W \sim \epsilon^2$ and $V \sim \epsilon^4$. We can now understand why it was sufficient to expand the covariant derivatives to first order in $\delta x$, $\delta z^j$ and $\delta \zb^j$. It is easy to check that higher order terms would be subleading both in the covariant derivatives as well as in $V$. 
For what follows it will be useful to write the expressions \eqref{eq:linearepcovu} and \eqref{eq:linearepcovz} more compactly using the notation:
\begin{equation}
\label{eq:compactcovmulti}
D_I W = (A_{Ij}+B_{Ij}\Delta y)\delta z^{j}+C_{Ij}\delta\bar{z}^{j}+G_{I}\delta x+\eta_{I}\Delta y.
\end{equation}
Here the index $I$ runs over $u$ and all $z^i$, where $I=0$ is identified with $u$ and $I=i$ with $i=1, \ldots, n$ corresponds to $z^i$. A summation over the index $j$ is implied. While being simple, the notation \eqref{eq:compactcovmulti} obscures some of the structure evident in \eqref{eq:linearepcovu} and \eqref{eq:linearepcovz}. In particular, note that 
\begin{align}
\label{eq:restrict1} 
B_{0i} \ & = \pd_u \pd_{z^i} a =0 &\textrm{for } i=1, \ldots, n \ , \\
B_{ij} \ & = B_{ji} = \pd_{z^i} \pd_{z^j} a & \textrm{for } i,j=1, \ldots, n \ , \\
\label{eq:restrictlast} G_i \ & = 2 A_{0i} &\textrm{for } i=1, \ldots, n \ .
\end{align} 
In the following, it will be convenient to work with real fields only. Writing $z^i = v^i + i w^i$ and $\zb^i=v^i - i w^i$ we can rewrite \eqref{eq:compactcovmulti} in terms of the displacements $\delta v^j$ and $\delta w^j$:
\begin{equation}
\label{eq:compactcovmultireal}
D_I W = (A_{Ij} + C_{Ij}+B_{Ij}\Delta y)\delta v^{j} + i (A_{Ij} - C_{Ij} + B_{Ij}\Delta y)\delta w^{j} +G_{I}\delta x+\eta_{I}\Delta y \ .
\end{equation}

We are now in a position to write down the $F$-term potential at quadratic order in the displacements, starting from its definition,
\begin{equation}
\label{eq:fpot}
V_{F}=e^{\mathcal{K}} \mathcal{K}^{I\bar{J}}D_{I} W \overline{D_J W},
\end{equation}
and insert our expressions \eqref{eq:compactcovmultireal}. The resulting potential can be written as a quadratic form:
\begin{equation}
\label{eq:fpot2}
V_F = \frac{1}{2} \mathbf{\Delta}^{T}\mathbf{\mathcal{D}}(\Delta y)\mathbf{\Delta} + [\mathbf{b}(\Delta y, \eta_I)]^{T}\mathbf{\Delta}+\mu^{2}(\Delta y)^{2} \ ,
\end{equation}
whose individual terms we will now explain. For one, $\mathbf{\Delta}$ is a vector with $(2n+1)$ entries containing the displacements $\mathbf{\Delta}=(\delta x, \delta v^i, \delta w^i)^{T}$. 
Also, $\mu^{2}=e^{\mathcal{K}}\mathcal{K}^{I\bar{J}}\eta_{I}\bar{\eta}_{\bar{J}}$ is the squared mass of the naive inflaton potential. Furthermore, $\mathbf{\mathcal{D}}$ is the real symmetric $(2n+1) \times (2n+1)$ matrix of the second derivatives of the scalar potential with respect to the displacements $\delta x, \delta v^i, \delta w^i$. Explicitly, it is given by
\begin{equation}
\label{eq:d}
\mathbf{\mathcal{D}}= \begin{pmatrix}
\mathcal{D}_{xx} &  \mathcal{D}_{xv^j} & \mathcal{D}_{xw^j}\\
\mathcal{D}_{v^i x} & \mathcal{D}_{v^i v^j}& \mathcal{D}_{v^i w^j}\\
\mathcal{D}_{w^i x} & \mathcal{D}_{w^i v^j} & \mathcal{D}_{w^i w^j}  
\end{pmatrix} \ ,
\end{equation}
with:
\begin{align}
\label{eq:delements}
\mathcal{D}_{xx}&= 2 \ e^{\mathcal{K}} \mathcal{K}^{I\bar{J}}G_{I}\overline{G_{J}} \ , \\
\nonumber \mathcal{D}_{x v^i} = \mathcal{D}_{v^i x}&=e^{\mathcal{K}} \mathcal{K}^{I\bar{J}} \left[G_I \overline{(A_{Ji} + C_{Ji}+B_{Ji}\Delta y)} + (A_{Ii} + C_{Ii}+B_{Ii}\Delta y) \overline{G_J} \right] \ , \\
\nonumber \mathcal{D}_{x w^i} = \mathcal{D}_{w^i x}&=e^{\mathcal{K}} \mathcal{K}^{I\bar{J}} \left[ - i G_I \overline{(A_{Ji} - C_{Ji}+B_{Ji}\Delta y)} + i (A_{Ii} - C_{Ii}+B_{Ii}\Delta y) \overline{G_J} \right] \ , \\
\nonumber \mathcal{D}_{v^i v^j} = \mathcal{D}_{v^j v^i}&=e^{\mathcal{K}} \mathcal{K}^{I\bar{J}} \left[ (A_{Ii} + C_{Ii}+B_{Ii}\Delta y) \overline{(A_{Jj} + C_{Jj}+B_{Jj}\Delta y)} + \right. \\ \nn & \hphantom{AAAAl} \left. + (A_{Ij} + C_{Ij}+B_{Ij}\Delta y) \overline{(A_{Ji} + C_{Ji}+B_{Ji}\Delta y)} \right] \ , \\
\nonumber \mathcal{D}_{v^i w^j} = \mathcal{D}_{w^j v^i}&=e^{\mathcal{K}} \mathcal{K}^{I\bar{J}} \left[ -i (A_{Ii} + C_{Ii}+B_{Ii}\Delta y) \overline{(A_{Jj} - C_{Jj}+B_{Jj}\Delta y)} + \right. \\ \nn & \hphantom{AAAAl} \left. + i (A_{Ij} - C_{Ij}+B_{Ij}\Delta y) \overline{(A_{Ji} + C_{Ji}+B_{Ji}\Delta y)} \right] \ , \\
\nonumber \mathcal{D}_{w^i w^j} = \mathcal{D}_{w^j w^i}&=e^{\mathcal{K}} \mathcal{K}^{I\bar{J}} \left[ (A_{Ii} - C_{Ii}+B_{Ii}\Delta y) \overline{(A_{Jj} - C_{Jj}+B_{Jj}\Delta y)} + \right. \\ \nn & \hphantom{AAAAA} \left. + (A_{Ij} - C_{Ij}+B_{Ij}\Delta y) \overline{(A_{Ji} - C_{Ji}+B_{Ji}\Delta y)} \right] \ .
\end{align}
The elements of the vector $\mathbf{b} = (b_x, b_{v^i}, b_{w^i})^T$ are given by the first derivatives of the $F$-term potential (evaluated at the minimum, i.e.~at $\mathbf{\Delta}=0$). Explicitly, we have
\begin{align}
\label{eq:bz}
b_{x} = {[\pd_{(\delta x)} V]}_{\star} &=e^{\mathcal{K}} \mathcal{K}^{I\bar{J}} \left[G_I \overline{\eta_J} + \eta_I \overline{G_J} \right] \Delta y \ , \\
\nn b_{v^i} = {[\pd_{(\delta v^i)} V]}_{\star} &=e^{\mathcal{K}} \mathcal{K}^{I\bar{J}} \left[(A_{Ii} + C_{Ii}+B_{Ii}\Delta y) \overline{\eta_J} + \eta_I \overline{(A_{Ji} + C_{Ji}+B_{Ji}\Delta y)} \right] \Delta y \ , \\
\nn b_{w^i} = {[\pd_{(\delta w^i)} V]}_{\star} &=e^{\mathcal{K}} \mathcal{K}^{I\bar{J}} \left[ i (A_{Ii} - C_{Ii}+B_{Ii}\Delta y) \overline{\eta_J} - i \eta_I \overline{(A_{Ji} - C_{Ji}+B_{Ji}\Delta y)} \right] \Delta y \ .
\end{align}
We can now determine the displacements $\delta x$, $\delta v^i$ and $\delta w^i$ as functions of $\Delta y$ by minimising the potential \eqref{eq:fpot2}. The unique minimum at each value of $\Delta y$ is found by solving
\begin{equation}
\mathbf{\mathcal{D}}\mathbf{\Delta}_{min}=- \mathbf{b} \quad.
\end{equation}
We find
\begin{equation}
\label{eq:min}
\Rightarrow \quad \mathbf{\Delta}_{min} =- \mathbf{\mathcal{D}}^{-1}\mathbf{b} =- \frac{\text{adj}[\mathbf{\mathcal{D}]}}{\text{det}[\mathbf{\mathcal{D}}]} \ \mathbf{b},
\end{equation}
where $\text{adj}[\mathbf{\mathcal{D}]}$ is the adjugate matrix of $\mathbf{\mathcal{D}}$. By substituting the solution $\mathbf{\Delta}_{min}$ back into \eqref{eq:fpot2} we arrive at the effective potential 
\begin{equation}
\label{eq:vinf}
V_{eff}(\Delta y)= -\frac{1}{2} \mathbf{b}^{T}(\Delta y)\mathcal{D}^{-1}(\Delta y) \mathbf{b}(\Delta y)+\mu^{2}\Delta y^{2}.
\end{equation}
This is the main result of this section. We have derived an expression for the effective potential with backreaction taken into account, i.e.~$V_{eff}$ is the potential along the flattest trajectory away from the SUSY minimum. Note that it still remains to be checked whether this potential is suitable for inflation. Further, recall that the above is only valid as long as backreaction of complex structure moduli is weak, such that terms cubic in $\delta x$ etc.~can be ignored. In the following section we will show that this can be achieved by tuning all $\eta_I$ small. 

However, before analysing \eqref{eq:vinf} further we can already make the following observation: even if backreaction is under control (i.e.~the displacements $\delta x$ etc.~are small) the effect of backreaction onto the inflaton potential is not negligible. Without backreaction the potential would be just given by $\mu^2 (\Delta y)^2= e^{\mathcal{K}} \mathcal{K}^{I \bar{J}} \eta_I \overline{\eta_J} (\Delta y)^2$, which is quadratic in the small quantities $\eta_I$. Note that all entries of the vector $\mathbf{b}$ \eqref{eq:bz} are linear in the small quantities $\eta_I$, while $\mathcal{D}$ does not depend on $\eta_I$ at all. As a result, the first term in \eqref{eq:vinf} containing the effects of backreaction is quadratic in $\eta_I$. As there are no other small parameters in our setup we find that the first term in \eqref{eq:vinf} is not parametrically suppressed w.r.t.~the naive inflaton potential. On the contrary, both terms in \eqref{eq:vinf} are equally important and the effective potential can differ significantly from the naive inflaton potential.

In the next section, we will analyse the effective potential in more detail. In particular, we will find:
\begin{itemize}
\item For small and intermediate $\Delta y$ the effective potential does in general not behave like a simple monomial in $\Delta y$. While the naive inflaton potential is quadratic by construction, backreaction will change this behaviour for intermediate $\Delta y$.
\item However, for large enough $\Delta y$ the effective potential can again be approximated by a parabola $V_{eff} = \mu_{eff}^2 (\Delta y)^2$. We are thus left with a sizable interval in field space where the effective potential is essentially quadratic. Thus it is in principle suitable for realising quadratic large field inflation.
\end{itemize}

\subsection{Quantifying backreaction}
\label{sec:backreaction}
In this section we wish to determine $\mathbf{\Delta}_{min} (\Delta y)$ and check that backreaction can indeed be controlled. By substituting $\mathbf{\Delta}_{min} (\Delta y)$ into \eqref{eq:fpot2} we will also be able to study the effective potential as a function of $\Delta y$.

To perform the next steps analytically and in full generality is not practical. The inverse matrix $\mathcal{D}^{-1}$ and thus $\mathbf{\Delta}_{min}$ will typically be complicated expressions in the parameters $A_{Ii}$, $B_{Ii}$, $C_{Ii}$, $G_I$ and $\eta_I$, which will obscure the points we wish to make in this section. 

To circumvent these complications, one can study backreaction and the effective potential numerically, and we will do so in section \ref{sec:brnum}. Here we adopt a different approach. In particular, we wish to show that by tuning $\eta_I$ small backreaction of complex structure moduli can be controlled. For this analysis the exact numerical values of the parameters $A_{Ii}$, $B_{Ii}$, $C_{Ii}$ and $G_I$ as well as $\mathcal{K}^{I \bar{J}}$ are not important; all we need to know is that they are not tuned small. Thus, to simplify the following calculations, we assume
\begin{align}
|A_{Ii}| \sim |B_{Ii}| \sim |C_{Ii}| \sim |G_{I}| \sim \mathcal{K}^{I \bar{J}} & \sim \O(1) \ , \\
|\eta_I |& \sim \epsilon^2 \ll 1 \ .
\end{align}
Then the matrix $\mathbf{\mathcal{D}}$ and the vector $\mathbf{b}$ are given by:
\begin{equation}
\label{eq:matrix1}
\mathbf{\mathcal{D}}= e^{\mathcal{K}}
\begin{pmatrix}
\O(1) & \O(1)+ \O(1)\Delta y & \hdots & \O(1) + \O(1)\Delta y\\
\O(1)+\O(1)\Delta y \hphantom{A}& (\O(1)+\O(1)\Delta y)^2 & \hdots & (\O(1)+\O(1)\Delta y)^{2}\\
\vdots & \vdots & \ddots & \vdots& \\
\O(1)+\O(1)\Delta y\hphantom{A} & (\O(1)+\O(1)\Delta y)^{2} & \hdots & (\O(1)+\O(1)\Delta y)^2
\end{pmatrix},
\end{equation}
\begin{equation}
\label{eq:b}
\mathbf{b}= e^{\mathcal{K}}
\begin{pmatrix}
\O(1)\\
\O(1)+ \O(1)\Delta y \\
\vdots \\
\O(1)+ \O(1)\Delta y
\end{pmatrix}\epsilon^2 \ \Delta y.
\end{equation}
It is now straightforward to determine the dependence of ${\mathcal{D}}^{-1}$ on $\Delta y$. Recall that for a geometry with $n+1$ complex structure moduli $\mathcal{D}$ is a $(2n+1)\times (2n+1)$ matrix. Then one obtains:
\begin{equation}
\label{eq:invD1}
{\mathcal{D}}^{-1}= \frac{e^{-\mathcal{K}}}{\textrm{pol}^{4n} (\Delta y)}
\begin{pmatrix}
\textrm{pol}^{4n} (\Delta y) & \textrm{pol}^{4n-1} (\Delta y) & \hdots & \textrm{pol}^{4n-1} (\Delta y) \\
\textrm{pol}^{4n-1} (\Delta y) \hphantom{A}& \textrm{pol}^{4n-2} (\Delta y) & \hdots & \textrm{pol}^{4n-2} (\Delta y) \\
\vdots & \vdots & \ddots & \vdots& \\
\textrm{pol}^{4n-1} (\Delta y) \hphantom{A} & \textrm{pol}^{4n-2} (\Delta y) & \hdots & \textrm{pol}^{4n-2} (\Delta y)
\end{pmatrix},
\end{equation}
where $\textrm{pol}^d( \Delta y)$ symbolises a polynomial of degree $d$ in $\Delta y$. More precisely, $\textrm{pol}^d( \Delta y) = \sum_{m=0}^d p_{m} (\Delta y)^m$ with coefficients $p_m$ which depend on $A_{Ii}$, $B_{Ii}$, $C_{Ii}$, $G_I$ and $\mathcal{K}^{I \bar{J}}$.

To arrive at \eqref{eq:invD1} we had to rely on several assumptions. For one, to be able to invert $\mathcal{D}$ it has to be non-degenerate. In addition, if $\mathcal{D}$ has a non-trivial substructure, it is certainly possible that there are cancellations when calculating the determinant and adjugate of $\mathcal{D}$. Then the polynomials appearing in ${\mathcal{D}}^{-1}$ would be of a lower degree than naively expected. We checked numerically that cancellations typically do not occur and hence it is justified to write $\mathcal{D}^{-1}$ as in \eqref{eq:invD1}.

We are now in a position to determine the displacements $\delta x$, $\delta v^i$ and $\delta w^i$ as functions of $\Delta y$:
\begin{equation}
\label{eq:deltamin}
\mathbf{\Delta}_{min}=
\begin{pmatrix}
\delta x \\
\delta v^i \\
\delta w^i
\end{pmatrix}_{min}
=
\begin{pmatrix}
\textrm{pol}^{4n} (\Delta y) \\
 \textrm{pol}^{4n-1} (\Delta y) \\
\textrm{pol}^{4n-1} (\Delta y)
\end{pmatrix}\frac{\epsilon^2 \ \Delta y}{\textrm{pol}^{4n} (\Delta y)},
\end{equation}
where in the above $\delta v^i$ and $\delta w^i$ represent all moduli of this type.

We can make the following observations. For one, the displacements $\delta x$, $\delta v^i$ and $\delta w^i$ are proportional to the small parameter $\epsilon^2$. Thus they are in principle small to the extent that $\epsilon^2$ is small. We used this fact in the previous section to neglect terms of the form $\epsilon \delta x$ etc.~in $D_I W$. However, given the expression \eqref{eq:deltamin} we can say much more about the dependence of $\delta x$, $\delta v^i$ and $\delta w^i$ on $\Delta y$. In particular, we can identify three regimes where the displacements behave differently:
\begin{enumerate}
\item $\Delta y \ll 1$: In this regime the polynomials in \eqref{eq:deltamin} will be dominated by their constant terms. It is then easy to see that $\delta x \sim \delta v^i \sim \delta w^i \sim \epsilon^2 \Delta y$. The displacements increase linearly with $\Delta y$, but they remain small in this regime. Backreaction is under control.
\item $\Delta y\sim O(1)$: no term in particular is expected to dominate in the polynomials of \eqref{eq:deltamin}. The displacements then behave as generic functions of $\Delta y$, possibly with regions of positive and negative slope. While the displacements are still suppressed by $\epsilon^2$, they can get enhanced in this regime if the term in the denominator of \eqref{eq:deltamin} (i.e.~the determinant of $\mathcal{D}$) becomes small. In this case backreaction is not completely under control and higher order terms in $\delta x$ etc.~cannot always be ignored.
\item $\Delta y\gg 1$: here the polynomials are dominated by the monomial with the highest degree: $\textrm{pol}^d (\Delta y) \sim (\Delta y)^d$. We then find the following: $\delta v^i, \delta w^i$ approach a constant, while $\delta x$ increases linearly with $\Delta y$. In particular, $\delta v^i \sim \delta w^i \sim \O(1) \epsilon^2$ while $\delta x \sim O(1) \epsilon^2 \Delta y$. The most dangerous modulus in this regime is then $\delta x$, as it increases linearly with $\Delta y$.
We can ignore higher order corrections in $\delta x$ to the potential as long as $\delta x \ll 1$, which requires $\Delta y \ll 1/\epsilon^2$. This condition is automatically satisfied as we are working under the assumption $0<\Delta y\ll1/{\epsilon}$. Therefore in this regime higher order corrections in $\delta x$ are negligible.
\end{enumerate}
In quadratic inflation one is interested in the regime of large displacements along the inflationary direction. As we have just shown, in this particular regime backreaction is completely under control up to maximal distances $\sim O(1/\epsilon)$. The parameter $\epsilon$ cannot be set to any arbitrary value, as this will affect both the phenomenology of inflation as well as the severity of tuning in the landscape. We will discuss this more thoroughly in section \ref{sec:landscape}. Let us here anticipate that it is feasible to have $(\Delta y)_{max}\sim O(10^{2})$ in units of the Planck mass. The important point is that there exist a regime of large field displacements where our assumptions about backreaction are justified. Therefore in this regime the approximation of the potential to quadratic order in $\delta x$, $\delta z^i$ and $\delta \bar{z}^i$ is valid. 

We now turn to the effective potential, which we already encountered in \eqref{eq:vinf}:
\begin{equation}
\nn V_{eff}= -\frac{1}{2}\mathbf{b}^{T}\mathcal{D}^{-1} \mathbf{b}+\mu^{2}\Delta y^{2} \ .
\end{equation}
In the previous section we already observed that both terms scale as $\epsilon^4$ and thus backreaction is not negligible. Here we will study its dependence on $\Delta y$.

Many observations from our analysis of the $\Delta y$-dependence of $\mathbf{\Delta}_{min}$ also apply here. We will be particularly interested in the regime $1 \ll \Delta y \ll 1 / {\epsilon}$. As we just argued, our expansion of the potential to second order is a good approximation of the $F$-term scalar potential (\ref{eq:fpot}) in this regime. In this region of field space, the inverse matrix $\mathbf{\mathcal{D}}^{-1}$ and the vector $\mathbf{b}$ are easy to write down:
\begin{align}
\label{eq:invmatrix}
\mathbf{\mathcal{D}}^{-1} & \simeq  e^{-\mathcal{K}}
\begin{pmatrix}
\O(1) & \O(1)\Delta y^{-1} & \hdots & \O(1)\Delta y^{-1}\\
\O(1)\Delta y^{-1} & \O(1)\Delta y^{-2} & \hdots & \O(1)\Delta y^{-2}\\
\vdots & \vdots & \ddots & \vdots \\
\O(1)\Delta y^{-1} & \O(1)\Delta y^{-2} & \hdots& \O(1)\Delta y^{-2}
\end{pmatrix},\\
\nonumber \quad\\
\label{eq:b2}
b &\simeq \ e^{\mathcal{K}}
\begin{pmatrix}
\O(1)\ \epsilon^2 \ \Delta y\\
\O(1)\ \epsilon^2 \ \Delta y^{2}\\
\vdots \\
\O(1)\ \epsilon^2 \ \Delta y^{2}
\end{pmatrix}.
\end{align}
In the regime of large $\Delta y$ the effective potential is then given by inserting the two above expressions \eqref{eq:invmatrix} and \eqref{eq:b2} into \eqref{eq:vinf}:
\begin{equation}
\label{eq:vmin2}
V_{eff}\simeq \Big(-O(1)e^{\mathcal{K}}\epsilon^{4}+\mu^{2}\Big)\Delta y^{2}\equiv \mu_{eff}^{2}\Delta y^{2},
\end{equation}
where $\mu^{2} = e^{\mathcal{K}} K^{I \bar{J}} \eta_I \overline{\eta_J} \sim e^{\mathcal{K}} \epsilon^{4}$. Some comments are in order. First, we find that for large $\Delta y$ the effective potential is a sum of two terms quadratic in $\Delta y$. The first one is due to backreaction on $\delta x, \delta z^i,\delta\bar{z}^i$ as one moves along $\Delta y$. The second term is the naive $\Delta y$ potential. The computation that we performed shows that those two contributions are of the same order of magnitude. Therefore we observe that, even though backreaction is under control in the regime under consideration, its effect on the potential is certainly not negligible. 

Secondly and most importantly, in the regime $1 \ll \Delta y \ll 1/\epsilon$ the potential is well approximated by a positive quadratic function. It is therefore in principle suitable for realising quadratic inflation. Notice however that the effective mass $\mu_{eff}$ is numerically smaller than the naive mass $\mu$.

Our result can be compared to previous studies of backreaction in axion monodromy inflation. In \cite{10114521, 14053652} it was found that backreaction of the inflaton potential on heavier moduli can flatten the inflaton potential at large field values. To be specific, for models of inflation with $\varphi^p$-potentials this can manifest itself in the reduction of the power $p$ at large field values. In our case we do not observe a reduction in the power $p$: our inflaton potential is quadratic for both small and large inflaton field values and flattening reduces the inflaton mass instead. This particular manifestation of flattening is a direct consequence of the mathematical structure of the supergravity scalar potential once we implement all the tuning conditions. Most importantly, the flattening we observe has the same physical origin as the effect described by \cite{10114521, 14053652}: it arises from integrating out heavier moduli.


By canonically normalising the inflaton we can then also determine the physical inflaton mass. Note that the inflaton direction is mainly given by $y$: at large $\Delta y$ the moduli $z^i$ are essentially fixed and $\delta x \sim \epsilon \Delta y$ only varies weakly with $y$. Thus, to leading order we can identify the inflaton with $\Delta y$. The effective Lagrangian for $\Delta y$ reads:
\begin{equation}
\mathcal{L}_{eff}=\mathcal{K}_{uu}(\partial \Delta y)^2-V_{eff}(\Delta y)=\mathcal{K}_{uu}(\partial \Delta y)^2-\mu^{2}_{eff}(\Delta y)^{2}.
\end{equation}
Therefore, at leading order the inflaton is simply obtained via the rescaling $\varphi=\sqrt{2\mathcal{K}_{uu}}\Delta y$ and the inflaton mass is given by $m_{\theta}^{2}=\mu^{2}_{eff}/\mathcal{K}_{uu}$. The constraint $\Delta y \ll 1 / {\epsilon}$ can now be translated into a constraint on the maximal initial displacement of $\varphi$. The field range of the inflaton is limited to $\varphi \ll \sqrt{2\mathcal{K}_{uu}} / \epsilon$.


This section can thus be summarised as follows: by tuning small $n+1$ parameters $a, \pd_{z^1} a, \dots, \pd_{z^n} a$, we can ensure that there exists a large range in field space in which backreaction is under control and the inflationary potential is in principle suitable for quadratic inflation. 


\subsection{Backreaction for less severe tuning}
\label{sec:tuneless}
In the previous sections we showed that by tuning $|a_i + \K_i a| \sim \epsilon^2$ (recall that $\epsilon \equiv |a|$) we can arrive at a potential for $\Delta y$ which is in principle suitable for inflation. Here, we wish to analyse whether backreaction can also be controlled for a less severe tuning. In particular, we will somewhat relax the tuning of $a_i$ and only require $|a_i| \sim |a| = \epsilon$, such that now $|a_i + \K_i a| \sim \epsilon$. We will argue that in this case we can still find an extended region in field space, where the inflaton potential is quadratic. In contrast to the previous sections, this regime will arise for $\Delta y \gg 1/ \epsilon$. As we will point out later, the backreaction of K\"ahler moduli cannot be neglected in this case, but inflation is still possible (as we will explain in section \ref{sec:kaehlermoduli}). Here we begin by analysing the backreaction of complex structure moduli. 

We start with the  covariant derivatives $D_u W$ and $D_{z^i} W$ expanded around $u_{\star}$ and $z_{\star}$ to first order in the displacements $\delta x$, $\delta z ^i$ and $\delta \zb^i$ as given in \eqref{eq:linearcovu} and \eqref{eq:linearcovz}. In what follows it will be most instructive to only work with two complex structure moduli $u$ and $z$. The analysis can be straightforwardly generalised to situations with further complex structure moduli. The expressions \eqref{eq:linearcovu} and \eqref{eq:linearcovz} can be written as
\begin{align}
\label{eq:267}D_u W &= (A_u + B_u \Delta y) \delta z + (C_u + F_u \Delta y) \delta \zb + (G_u + H_u \Delta y) \delta x + \eta_u \Delta y\ , \\
\label{eq:268}D_z W &= (A_z + B_z \Delta y) \delta z + (C_z + F_z \Delta y) \delta \zb + (G_z + H_z \Delta y) \delta x + \eta_z \Delta y\ .
\end{align}
The parameters $A_{u,z}$, $B_{u,z}$, $C_{u,z}$, $F_{u,z}$, $G_{u,z}$, $H_{u,z}$ and $\eta_{u,z}$ can simply be read off from \eqref{eq:linearcovu} and \eqref{eq:linearcovz}.
Due to the appearance of $a$ and $a_i$ in the above parameters there are hierarchies between the different terms in \eqref{eq:267} and \eqref{eq:268}. To keep track of this it will be convenient to rewrite as 
\begin{align}
\label{eq:DuWa} D_u W &= (\hat{A}_u + \epsilon \hat{B}_u \Delta y) \delta z + (\hat{C}_u + \epsilon \hat{F}_u \Delta y) \delta \zb + (\hat{G}_u +  \epsilon \hat{H}_u \Delta y) \delta x + \epsilon^2 \hat{\eta}_u \Delta y \ , \\
\label{eq:DzWa}D_z W &= (\hat{A}_z + \hat{B}_z \Delta y) \delta z + (\hat{C}_z + \epsilon \hat{F}_z \Delta y) \delta \zb + (\hat{G}_z + \epsilon \hat{H}_z \Delta y) \delta x + \epsilon \hat{\eta}_z \Delta y \ ,
\end{align}
where $\epsilon = |a|$. The hatted parameters then do not contain any small quantities and we will assume
\be
\hat{A}_{u,z} \sim \hat{B}_{u,z} \sim \hat{C}_{u,z} \sim \hat{F}_{u,z} \sim \hat{G}_{u,z} \sim \hat{H}_{u,z} \sim \hat{\eta}_{u,z} \sim \O (1) \ .
\ee
Here we will be exclusively interested in the region $\Delta y \gg 1/ \epsilon$. In this case terms like $A_u \delta z$ are subleading compared to $\epsilon B_u \Delta y \delta z$ etc. Suppressing subleading terms we can write
\begin{align}
\label{eq:DuWb} D_u W &= \left[ \epsilon \hat{B}_u \delta z +\epsilon \hat{F}_u \delta \zb + \epsilon \hat{H}_u \delta x + \epsilon^2 \hat{\eta}_u \right] \Delta y \ , \\
\label{eq:DzWb} D_z W &= \left[ \hat{B}_z \delta z + \epsilon \hat{F}_z \delta \zb + \epsilon \hat{H}_z \delta x + \epsilon \hat{\eta}_z \right] \Delta y \ .
\end{align}
We will now examine the backreaction on complex structure moduli. We assume that $\delta z \sim \delta x \sim \epsilon$, which we will confirm at the end. We can make the following observations.
\begin{enumerate}
\item At leading order in $\epsilon$ the potential is given by 
\be
V= e^{\K} \K^{z \zb} |D_z W|^2 = e^{\K} \K^{z \zb} |\hat{B}_z \delta z + \epsilon \hat{\eta}_z |^2 (\Delta y)^2 + \mathcal{O}(\epsilon^3) \ .
\ee
All contributions from $D_u W$ are strictly subleading.
\item The observation now is that the term $\hat{B}_z \delta z$ has enough freedom to cancel the term $\epsilon \hat{\eta}_z $ in $V$. As a result, backreaction of $z$ cancels the leading order inflaton potential completely. We hence find that the potential is minimised if the modulus $z$ is shifted at leading order as
\be
\label{eq:deltazone}
\delta z = \delta z_1 \equiv - \epsilon \frac{\hat{\eta}_z}{\hat{B}_z} \ .
\ee
The displacement of $x$ is left undetermined so far.
\end{enumerate}

As the potential vanishes at order $\epsilon^2$, we need to go beyond leading order. To this end we write 
\be
\delta z = \delta z_1 + \delta z_2 \ ,
\ee
where $\delta z_1$ was defined in \eqref{eq:deltazone}. We assume that $\delta z_2 \sim \epsilon^2$, which again will be justified \emph{a posteriori}. We insert $\delta z = \delta z_1 + \delta z_2$ into our expressions for $D_{u,z}W$ and keep the leading terms, which are now of order $\epsilon^2$. However, to collect all terms of order $\epsilon^2$ it is not sufficient to expand $D_{u,z}W$ only to linear order in $\delta z$, $\delta \zb$ and $\delta x$. Terms quadratic in $\delta z$ etc.~are now important. One can check explicitly that (cf.~\eqref{eq:covariant}) only one such term is of order $\epsilon^2$, while all other terms are suppressed further: the term in question is $\hat{L}_{zz} (\delta z)^2 \equiv \frac{i}{2} \left[a_{zzz} + \K_z a_{zz} \right] \Delta y (\delta z)^2$ in $D_{z} W$. Note that $\hat{L}_{zz}$ does not contain the small quantities $a$, $a_z$ or $\K_u$ and thus we take $\hat{L}_{zz} \sim \mathcal{O}(1)$. Then $\hat{L}_{zz} (\delta z_1)^2 \sim \epsilon^2$ and we need to include it in our expansion of $D_z W$. We thus have
\begin{align}
\label{eq:DuWc} D_u W &= \left[ \epsilon \hat{B}_u \delta z_1 +\epsilon \hat{F}_u \delta \zb_1 + \epsilon \hat{H}_u \delta x + \epsilon^2 \hat{\eta}_u \right] \Delta y + \O (\epsilon^3 \Delta y) \ , \\
\label{eq:DzWc} D_z W &= \left[ \hat{B}_z \delta z_2 + \epsilon \hat{F}_z \delta \zb_1 + \epsilon \hat{H}_z \delta x + \hat{L}_{zz} (\delta z_1)^2 \right] \Delta y + \O (\epsilon^3 \Delta y) \ .
\end{align}
It will now be convenient to write $\delta z_2=\epsilon^2 \delta \hat{z}_2$ and $\delta x = \epsilon \delta \hat{x}$ leading to
\begin{align}
\label{eq:DuWd} D_u W &= \left[ - \hat{B}_u \frac{\hat{\eta}_z}{\hat{B}_z} - \hat{F}_u \frac{\hat{\bar{\eta}}_z}{\hat{\bar{B}}_z} + \hat{H}_u \delta \hat{x} + \hat{\eta}_u \right] \epsilon^2 \Delta y + \O (\epsilon^3 \Delta y) \ , \\
\label{eq:DzWd} D_z W &= \left[ \hat{B}_z \delta \hat{z}_2 - \hat{F}_z \frac{\hat{\bar{\eta}}_z}{\hat{\bar{B}}_z} + \hat{H}_z \delta \hat{x} + \hat{L}_{zz} \frac{\hat{\eta}_z^2}{\hat{B}_z^2} \right] \epsilon^2 \Delta y + \O (\epsilon^3 \Delta y) \ ,
\end{align}
where we also used \eqref{eq:deltazone}. To determine $\delta \hat{z}_2$ and $\delta \hat{x}$ we examine how $V$ is minimised. So far we found that $V$ vanishes at order $\epsilon^2$ (and, automatically, also at order $\epsilon^3$) once backreaction is taken into account. Next we will show that such a cancellation does not in general occur at order $\epsilon^4$. To this end we write
\be
V= e^{\K} \K^{I \bar{J}} D_I W \overline{D_J W} = e^{\K} \left[ \K^{I \bar{J}} v_I \overline{v_J} \right] \epsilon^4 (\Delta y)^2 + \mathcal{O}(\epsilon^5 (\Delta y)^2) \ ,
\ee
where the indices $I,J$ run over the moduli $u,z$ and we defined
\begin{align}
v_u & \equiv - \hat{B}_u \frac{\hat{\eta}_z}{\hat{B}_z} - \hat{F}_u \frac{\hat{\bar{\eta}}_z}{\hat{\bar{B}}_z} + \hat{H}_u \delta \hat{x} + \hat{\eta}_u \ , \\
\label{eq:DzWd} v_z & \equiv \hat{B}_z \delta \hat{z}_2 - \hat{F}_z \frac{\hat{\bar{\eta}}_z}{\hat{\bar{B}}_z} + \hat{H}_z \delta \hat{x} + \hat{L}_{zz} \frac{\hat{\eta}_z^2}{\hat{B}_z^2} \ ,
\end{align}
We can now make the following observations.
\begin{enumerate}
\item The potential at order $\epsilon^4$ is a non-degenerate Hermitian inner product of the complex vector $v$ with itself. Thus, by construction it only vanishes if both $v_u$ and $v_z$ are zero. 
\item Note that the vector $v$ contains two complex, i.e.~four real components. However, $\delta \hat{z}_2$ and $\delta \hat{x}$ only contain three independent real degrees of freedom. By adjusting $\delta \hat{z}_2$ and $\delta \hat{x}$ it will thus not be possible in general to set $v_u=v_z=0$. 
\item Thus, at order $\epsilon^4$ the potential does not vanish in general once backreaction is taken into account. The leading contribution to the effective potential including backreaction is thus of the form
\be
V_{eff} = \mu_{eff}^2 (\Delta y)^2 \sim e^{\K} |\epsilon|^4 (\Delta y)^2 \ .
\ee
This is conclusion is valid for $\Delta y \gg 1/ \epsilon$.
\item Last, we confirm our assumptions regarding the size of the displacements. While $\delta \hat{z}_2$ and $\delta \hat{x}$ cannot cancel the potential at order $\epsilon^4$, they adjust such that the potential is minimised. In particular, they take values such that the Hermitian inner product $\K^{I \bar{J}} v_I \overline{v_J}$ is minimal. However, as $v_u$ and $v_z$ only contain parameters of size $\mathcal{O}(1)$, we can conclude that in general $\delta \hat{z}_2 \sim \mathcal{O}(1)$ and $\delta \hat{x} \sim \mathcal{O}(1)$. Thus we have
\be
\delta z_1 = - \epsilon \frac{\hat{\eta}_z}{\hat{B}_z} \sim \epsilon \ , \quad \delta z_2 = \epsilon^2 \delta \hat{z}_2 \sim \epsilon^2 \ , \quad \delta x = \epsilon \delta \hat{x} \sim \epsilon \ ,
\ee
as claimed at the beginning of this section.
\end{enumerate}

To summarise, in this section we observed that tuning $|a_z| \sim |a| \sim \epsilon$ is enough to ensure that there is a large interval in field space, where the potential including backreaction is quadratic. We find that for $\Delta y \gg 1/ \epsilon$ the F-term potential takes the form $V_{eff} = \mu_{eff}^2 (\Delta y)^2 + V_{LVS}$ with $\mu_{eff}^2 \sim e^{\K} |\epsilon|^4$. Furthermore, we find that $x$ as well as $z$ are only displaced by a small amount from their value at the global minimum: $\delta z \sim \delta \zb \sim \delta x \sim \epsilon$.

However, note that for large displacements $\Delta y \gg 1/ \epsilon$ the superpotential $W=w+au$ is dominated by $au \sim \epsilon \Delta y$ and thus evolves when $\Delta y$ is changing. As a result, the backreaction of K\"ahler moduli cannot be neglected in this case. Stabilisation according to the Large Volume Scenario fixes the volume as $\mathcal{V} \propto |W|$ and thus the volume will necessarily change when $\Delta y$ is evolving. Strictly speaking, the inflaton will not simply be given by $\Delta y$, but necessarily also involve the volume. We will discuss the consequences in the next section.


\subsection{K\"ahler moduli and backreaction}
\label{sec:kaehlermoduli}

In this section we briefly comment on the consequences of large displacements of $\Delta y$ for K\" ahler moduli stabilisation. The discussion is based on moduli stabilisation according to the LVS \cite{0502058}. In this framework, complex structures moduli are integrated out and give rise to a constant tree level superpotential $W$. The scalar potential for the K\" ahler moduli arises through the interplay of $\alpha^{'}$-corrections in the K\" ahler potential and non-perturbative corrections in the superpotential. This effective potential for K\" ahler moduli admits a non-supersymmetric AdS minimum at exponentially large volume:
\begin{equation}
\label{eq:volume}
\mathcal{V}\propto |W|e^{2\pi\tau_{s}},
\end{equation}
where $\tau_{s}$ is the real part of the K\" ahler modulus of the small cycle. After minimisation, the LVS scalar potential behaves as $V_{LVS}\sim -|W|^{2}/\mathcal{V}^{3}$. 

In our setup the tree level superpotential is linear in one of the complex structure moduli, i.e. $W=w+au$. As long as $au\ll w$, the superpotential is approximately constant and the modulus $u$ does not play any role in the stabilisation of the volume. However, large $\Delta y$ displacements can make the linear term dominant with respect to $w$. In this case $W$, hence the volume according to \eqref{eq:volume}, runs with $\Delta y$. Thus the complex structure modulus $u$ can potentially interfere with the K\" ahler moduli, through the volume of the Calabi-Yau manifold.\footnote{The interplay between K\" ahler and complex structure moduli in complex structure moduli inflation has been also recently studied in \cite{14107522}. The authors consider a somewhat different scenario, based on a racetrack scalar potential for the K\" ahler moduli. They obtain constraints on the running of $W$ from the destabilisation of the volume. Given these conditions, they point out the difficulties associated with large field inflation in a model with one complex structure modulus.}  Moreover in this case, as we will show, the dominant contribution to the potential for $\Delta y$ comes from the LVS potential. Then our study of the complex structure F-term potential is not sufficient to establish whether the $\Delta y$ direction is suitable for realising quadratic inflation. 

In what follows we do not wish to perform a complete analysis of the issue that we have just presented. Rather, we would like to describe more specifically how this problem affects our work and suggest that inflation might nevertheless work.

Let us then separately discuss the two setups that were presented in sections \ref{sec:branalytical} and \ref{sec:tuneless} respectively. The first case, where $|a|\sim \epsilon$, $|a_{i}+\mathcal{K}_{i}a|\sim \epsilon^{2}$, is not affected by the discussion above. Indeed, it was assumed that the inflaton displacement is restricted to the region $\Delta y\ll 1/\epsilon$. In this regime we have $a\Delta y\ll w\sim O(1)$ and the superpotential is always dominated by the constant term.

\begin{figure}
	\centering
		\includegraphics[width=0.50\textwidth]{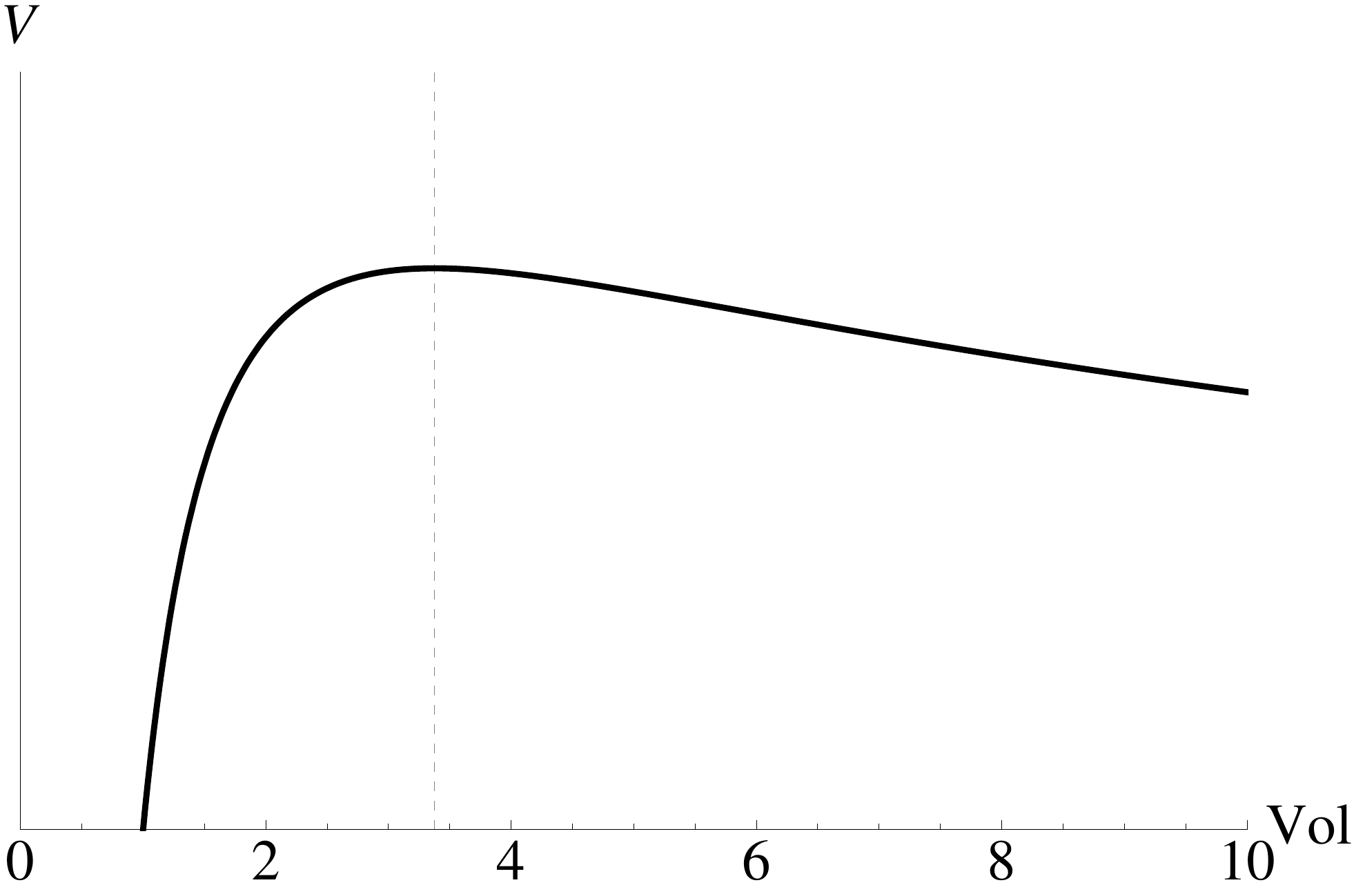}
	\caption{Total potential \eqref{eq:inflationarylvs} as a function of the volume $\mathcal{V}\sim e^{A}|a| \Delta y$. The dashed line intersects the potential at its maximum. Inflation could take place in the region on the left of the extremum. We normalised the $x$-axis such that $\langle \mathcal{V}\rangle=1$.}
\label{fig:kaehler}
\end{figure}

The second setup requires more attention. The complex structure moduli scalar potential is under explicit control only for $\Delta y\gg 1/\epsilon$. In this regime $a\Delta y\sim \epsilon \Delta y\gg w$, when $w\sim O(1)$. As we argued above, K\" ahler moduli stabilisation is certainly an important issue in this case. We focus on the relevance of the LVS potential for the candidate inflationary direction $\Delta y$. The starting point is the potential
\begin{equation}
\label{eq:vtot}
V_{tot}(\Delta y)=V_{eff}(\Delta y)+V_{LVS}(\Delta y)+V_{uplift}(\Delta y),
\end{equation}
where $V_{eff}\sim |\epsilon|^{4}(\Delta y)^{2}/\mathcal{V}^{2}$ is the effective potential computed in section \ref{sec:tuneless} and $V_{LVS}\sim |W|^{2}/\mathcal{V}^{3}$. We have also included a term to uplift to a dS vacuum. Notice that $V_{LVS}$ and $V_{uplift}$ depend on $\Delta y$ through $W$ and the volume, according to \eqref{eq:volume}. In particular, the effective potential $V_{eff}$ is suppressed with respect to $V_{LVS}$ by $\epsilon^{2}\mathcal{V}$, because $V_{LVS}\sim |W|^{2}/\mathcal{V}^{3}\sim \epsilon^{2}(\Delta y)^{2}/\mathcal{V}^{3}$ in the regime $\Delta y\gg 1/\epsilon$. In order to remain in the LVS framework, we tune $\epsilon$ such that $\epsilon^{2}\mathcal{V}\ll 1$. \footnote{Given a certain size of $\epsilon$, this bounds the volume $\mathcal{V}$. The limited size of $\mathcal{V}$ in large field models of this type has also been discussed in \citep{14043711} and plays a role in our Appendix A. A more general study of bounds on the volume has appeared in \citep{14114636} after the first version of this paper was submitted.} It is therefore clear that in this setup the relevant potential for $\Delta y$ comes from the interplay of the LVS and the uplift potentials, i.e.~$V_{tot}(\Delta y)\simeq V_{LVS}(\Delta y)+V_{uplift}(\Delta y)$. One can now perform a study of this potential, which necessarily depends on the functional form of the desired uplift. We focus on a scenario where the latter is provided by some hidden matter fields which develop non-vanishing VEVs through minimisation of their F- and D-term potentials \cite{12065237} (see also \cite{14091931} for a recent discussion). In this case the total scalar potential \eqref{eq:vtot}, neglecting $V_{eff}$, is given by \cite{14091931}:
\begin{equation}
\label{eq:vtotuplift}
V_{tot}(\mathcal{V})\propto \frac{e^{-4\pi\tau_{s}}}{\mathcal{V}}\Bigg[\mathcal{V}^{1/3}\delta-\sqrt{\ln\Big(\frac{\mathcal{V}}{W}\Big)}\Bigg],
\end{equation}
where $\delta$ is a numerical factor depending on the $U(1)$ charges of the big cycle modulus and the matter fields and \eqref{eq:volume} was used. At the minimum one imposes $\langle V_{tot}\rangle=0$ to achieve a Minkowski vacuum. Therefore at the minimum $\langle \mathcal{V}\rangle^{1/3}\delta=\ln\Big(\langle \mathcal{V}\rangle/|W|\Big)$. The total potential \eqref{eq:vtotuplift} can thus be rewritten as
\begin{equation}
\label{eq:inflationarylvs}
V_{tot}(\mathcal{V})\propto \frac{e^{-4\pi\tau_{s}}}{\mathcal{V}}\Big[\mathcal{V}^{1/3}-\langle \mathcal{V}\rangle^{1/3}\Big].
\end{equation}
This potential is monotonically rising from $0$ to $V_{max}=(3/2)^{3}\langle \mathcal{V}\rangle$, then decreases and vanishes asymptotically (see fig.~\ref{fig:kaehler}).

Since $\mathcal{V}\sim e^{2\pi\tau_{s}}|a|\Delta y$, the total potential rises monotonically as a function of $\Delta y$ up to $(\Delta y_{max}/y_{\star})\simeq 3.4/|a|$, where $y_{\star}$ is the value of the $y$ at the minimum. 
The inflationary range can be now found by canonically normalising $\Delta y$, i.e.~by defining $\varphi=\sqrt{\mathcal{K}_{uu}}\Delta y$. 
We conclude that for $\varphi\leq 3.4/(|a|x_{\star})$ the potential \eqref{eq:inflationarylvs} is monotonically rising. Notice that generically this is a sizable range, despite the fact that $x$ is stabilised in the LCS regime, as we have tuned $|a|$ small. 

The results of this section can be summarised as follows. We found that in the setup described in section \ref{sec:branalytical} the complex structure moduli do not affect K\" ahler moduli stabilisation. On the contrary, the setup described in section \ref{sec:tuneless} implies an interplay between the volume modulus and the inflationary direction $\Delta y$ in the regime of large field displacements. We found that in this case the LVS and uplift potentials give the dominant contribution to the total potential for $\Delta y$. By focusing on D-term uplifting from hidden sector matter fields, we showed that the total potential \eqref{eq:inflationarylvs} is still monotonically rising throughout a sizable range for $\Delta y$. As such, it might be suitable for realising large field inflation. Rather than focusing on a more detailed analysis of the potential, which is left for future work, we now provide some numerical examples of the effective inflationary potential including backreaction. 


\section{Numerical Examples}
\label{sec:brnum}
In this section we study the backreaction of moduli numerically. The examples presented in this section will be based on the analysis performed in sections \ref{sec:branalytical} and \ref{sec:backreaction}, where we tune $|a_i + \K_i a| \sim \epsilon^2$ with $|a|=\epsilon$. To be specific, we generate random values for coefficients in the scalar potential expanded to second order in $\delta z^i$, $\delta \zb^i$ and $\delta x$. We then determine $\delta z^i$, $\delta \zb^i$ and $\delta x$ which minimise the potential as a function of $\Delta y$ explicitly. In practice, it will be more convenient to work with the real and imaginary parts of $\delta z^i=\delta v^i+i\delta w^i$. Finally, we also determine the effective inflaton potential once backreaction is taken into account.

Before we present explicit examples a few words of warning are in order. For one, our examples do not arise from an explicit choice of geometry and flux numbers. Instead, we randomly generated parameters in the supergravity scalar potential. Hence it still needs to be checked whether the parameter values in our examples can arise for a given choice of geometry and fluxes (e.g.~along the lines of \cite{14097075}). In particular, the numerical examples we show only exhibit three or four complex structure moduli. In such a construction the number of available fluxes is also low and thus the ability to tune parameters in the scalar potential is severely restricted. Hence it is certainly possible that a model based on a choice of compactification geometry cannot reproduce the exact numerical data shown below. As long as one keeps this caveat in mind the following numerical examinations are nevertheless very instructive. The examples we show are not special in any way but rather exhibit the typical behaviour that we find for the supergravity models studied in this paper. In particular, we find that choosing different numerical values does not change the qualitative features significantly. Thus, while a realistic geometry and choice of fluxes might not be able to reproduce the following examples exactly, we are confident that such a realistic model will exhibit a similar behaviour. 

\subsection{Three-moduli-model}
Here we present a toy model with three complex structure moduli $u$, $z^1$ and $z^2$. We use this example to illustrate the analytical results from sections \ref{sec:branalytical} and  \ref{sec:backreaction}. As described there, to assess backreaction we expand the relevant part of the supergravity scalar potential $V= e^{\K} \K^{I \bar{J}} D_I W \overline{D_J W}$
to second order in 
\be
\nn \delta x = \delta \ \textrm{Re}(u) , \ \ \delta v^1= \delta \ \textrm{Re}(z^1) , \ \ \delta w^1= \delta \ \textrm{Im}(z^1), \ \ \delta v^2= \delta \  \textrm{Re}(z^2) \ \ \textrm{and} \ \ \delta w^2= \delta \ \textrm{Im}(z^2) 
\ee
about the global minimum. To this end we need to expand $D_I W$ to first order in $\delta x$, $\delta v^i$ and $\delta w^i$. The resulting expression can be parameterised as in \eqref{eq:compactcovmultireal}:
\begin{equation}
\label{eq:DIWparameters}
D_I W = (A_{Ij} + C_{Ij}+B_{Ij}\Delta y)\delta v^{j} + i (A_{Ij} - C_{Ij} + B_{Ij}\Delta y)\delta w^{j} +G_{I}\delta x+\eta_{I}\Delta y \ .
\end{equation}
To study the potential numerically, we will generate random values for the parameters appearing in \eqref{eq:DIWparameters}. However, not all parameters are completely unconstrained. As argued in the previous section we need to tune all $|\eta_I|$ small to control backreaction. In our numerical simulation we implement this as follows: we generate values for the parameters, such that 
\begin{align}
|A_{Ij} |, |B_{Ij} |, |C_{Ij} |, |G_I| \ & \sim \mathcal{O}(1) \ , \\
|\eta_I| \ & \sim \mathcal{O}(10^{-4})
\end{align}
Here we tune all $\eta_I$ equally small as described in section \ref{sec:branalytical}.
The explicit values $\mathcal{O}(1)$ and $\mathcal{O}(10^{-4})$ are not important -- the crucial point is the hierarchy between $|\eta_I|$ and the remaining parameters.\footnote{This is done in practice as follows: for both the real and imaginary parts of $A_{Ij} , B_{Ij} , C_{Ij} , G_I$ we generate uniformly distributed random numbers in the range $[-1.5, -0.5]$ or $[0.5, 1.5]$. The real and imaginary parts of $\eta_I$ are chosen from uniformly distributed random numbers in the range $[-1.5, -0.5] \cdot 10^{-4}$ or $[0.5, 1.5] \cdot 10^{-4}$.} In addition, we ensure that the values generated for the parameters in \eqref{eq:DIWparameters} also obey the relations \eqref{eq:restrict1} -- \eqref{eq:restrictlast}.

To arrive at a numerical expression for $V$ we will also require a numerical K\"ahler metric. This is generated as a complex $3 \times 3$ matrix whose entries are $| \K^{I \bar{J}}| \sim \mathcal{O}(1)$. We ensure that it is both hermitian and positive definite.\footnote{In practice we generate a $3 \times 3$ matrix $M$ with random complex entries of magnitude $\mathcal{O}(1)$, which we draw from uniformly distributed random numbers. The inverse K\"ahler metric is then obtained as $M^{\dagger} M$. This is positive semi-definite by construction.} In addition, there is the factor $e^{\K}$ which multiplies the whole potential. As it will only affect the overall scale of $V$ we set it to $e^{\K}=1$ for simplicity.

We begin by listing a choice of parameter values for our first numerical example. The inverse K\"ahler metric is given by
\be
\K^{I \bar{J}}
= \begin{pmatrix}
  \K^{u \bar{u}} & \K^{u \bar{z}^1} & \K^{u \bar{z}^2} \\
  \K^{z^1 \bar{u}} & \K^{z^1 \bar{z}^1} & \K^{z^1 \bar{z}^2} \\
  \K^{z^2 \bar{u}} & \K^{z^2 \bar{z}^1} & \K^{z^2 \bar{z}^1}
\end{pmatrix}
=\begin{pmatrix}
  1.085 & -0.714 - 0.539 \ i & -0.108 + 0.409 \ i \\
  -0.714 + 0.539 \ i & 1.133 & -0.192 - 0.634 \ i \\
  -0.108 - 0.409 \ i & -0.192 + 0.634 \ i & 0.854
 \end{pmatrix}
\ee
We further have
\vspace{\baselineskip}

\begin{tabular}{ l l l }
  $A_{01}=\hphantom{-} 1.146 + 0.939 \ i$ , & $A_{11}=-1.376 - 0.935 \ i$ , & $A_{21}=-1.316 - 0.604 \ i$ , \\
  $A_{02}=-0.515 - 1.399 \ i$ , & $A_{12}=-1.300 + 0.925 \ i$ , & $A_{22}=\hphantom{-} 0.958 - 1.251 \ i$ , \\
  $B_{01}=\hphantom{-} 0$ , & $B_{11}=\hphantom{-} 0.945 + 0.625 \ i$ , & $B_{21}=-0.919 - 1.418 \ i$ , \\
  $B_{02}=\hphantom{-} 0$ , & $B_{12}=-0.919 - 1.418 \ i$ , & $B_{22}=\hphantom{-} 0.650 + 1.026 \ i$ , \\
  $C_{01}=-1.010 - 1.094 \ i$ , & $C_{11}=\hphantom{-} 0.904 + 1.483 \ i$ , & $C_{21}=-1.057 - 0.690 \ i$ , \\
  $C_{02}=-1.369 - 0.953 \ i$ , & $C_{12}=-0.527 + 0.927 \ i$ , & $C_{22}=-0.647 - 1.460 \ i$ , \\
  $G_{0 \hphantom{1}}=-0.826 + 0.627 \ i$ , & $G_{1 \hphantom{1}}= \hphantom{-} 2.292 + 1.878 \ i$ , & $G_{2 \hphantom{1}}=-1.030 - 2.798 \ i$ , 
\end{tabular}
\vspace{\baselineskip}

\noindent as well as
\begin{align}
\nn \eta_{0 \hphantom{A}}&=(0.889+0.779 \ i) \cdot 10^{-4} \, \\
\nn \eta_{1 \hphantom{A}}&=(1.082-0.847 \ i) \cdot 10^{-4} \ , \\
\nn \eta_{2 \hphantom{A}}&=(0.725-1.472 \ i) \cdot 10^{-4} \ , 
\end{align}

\begin{figure}[t]
 \subfloat[][]{
 \includegraphics[width=0.46\textwidth]{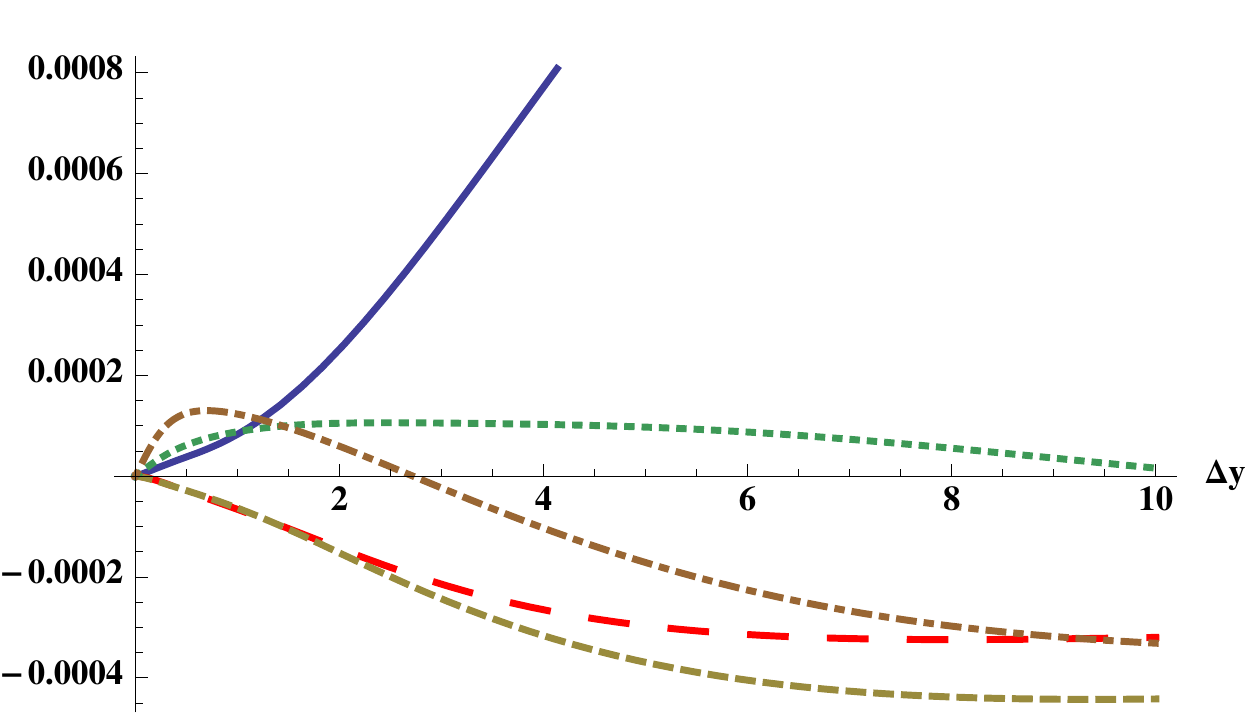}
 }
 \ \hspace{0.5mm} \hspace{5mm} \
 \subfloat[][]{
 \includegraphics[width=0.46\textwidth]{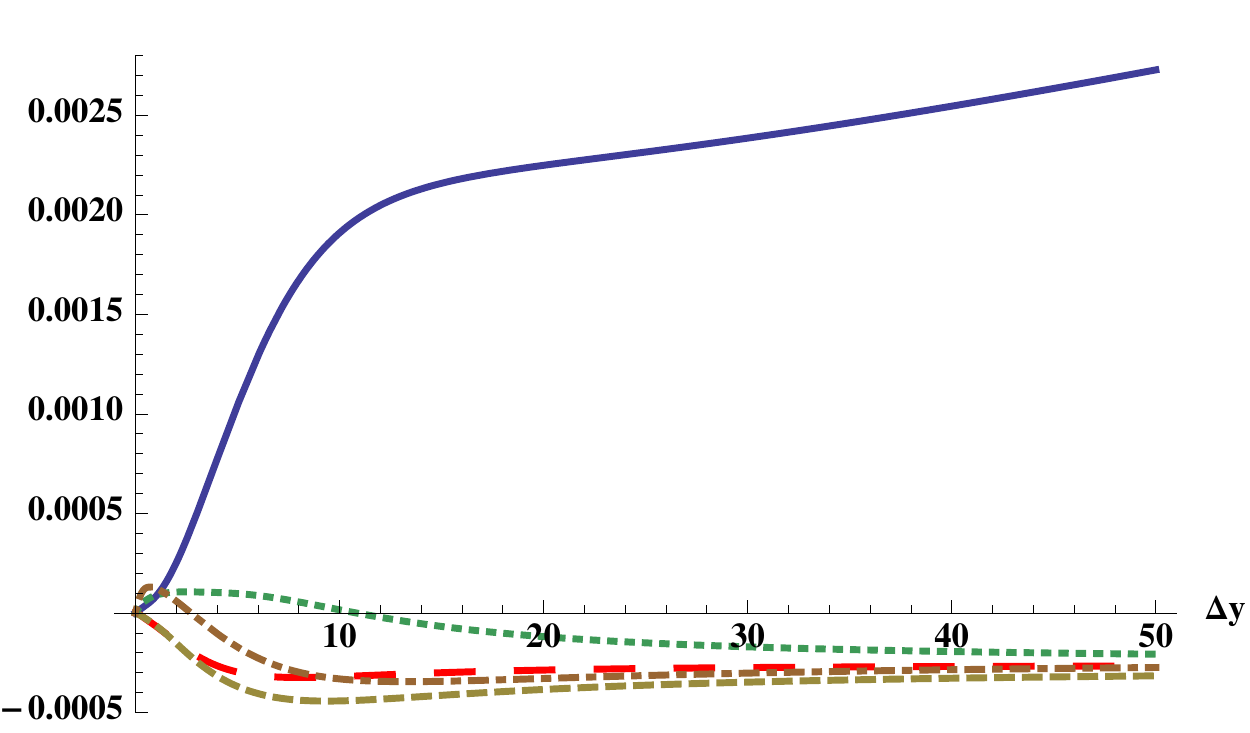}
 }
\caption{Plots of the displacements $\delta x$ (blue, solid), $\delta v^1$ (red, long dashes), $\delta w^1$ (ochre, short dashes), $\delta v^2$ (green, dotted) and $\delta w^2$ (brown, dot-dashed) vs.~$\Delta y$.}
\label{fig:n2xa}
\end{figure}

\begin{figure}[t]
 \subfloat[][]{
 \includegraphics[width=0.46\textwidth]{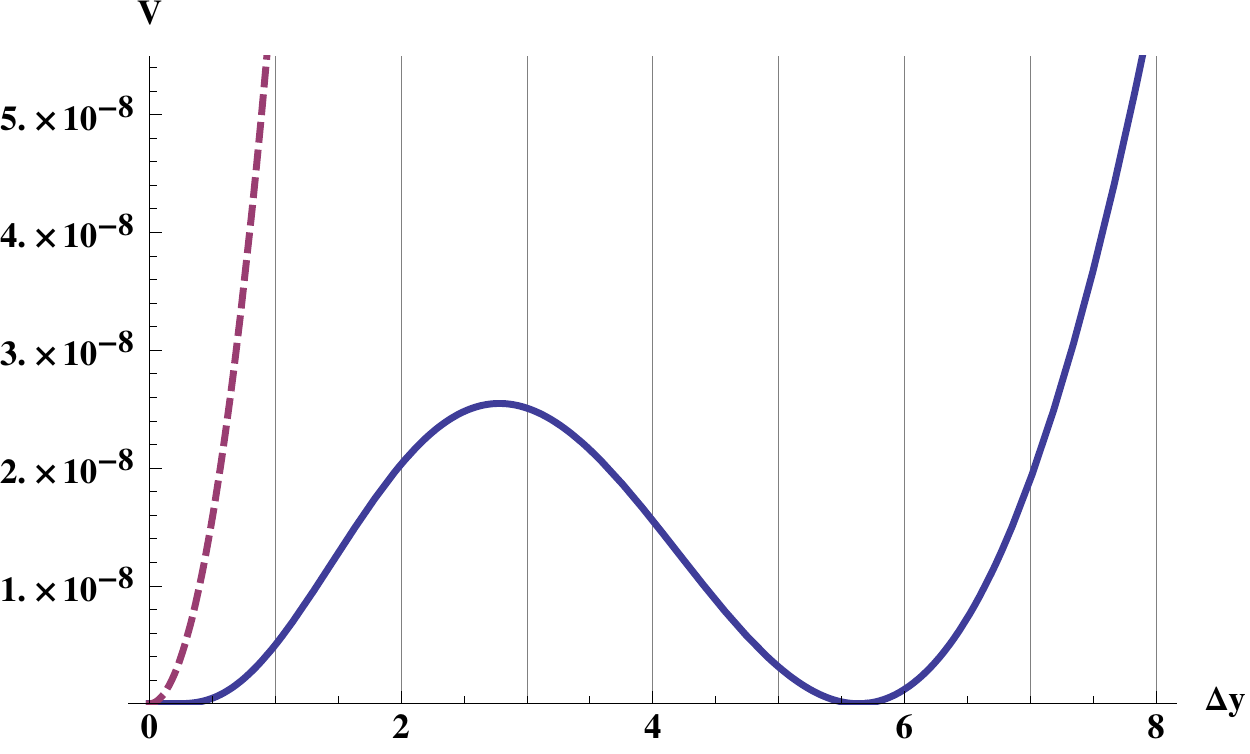}
 }
 \ \hspace{0.5mm} \hspace{5mm} \
 \subfloat[][]{
 \includegraphics[width=0.46\textwidth]{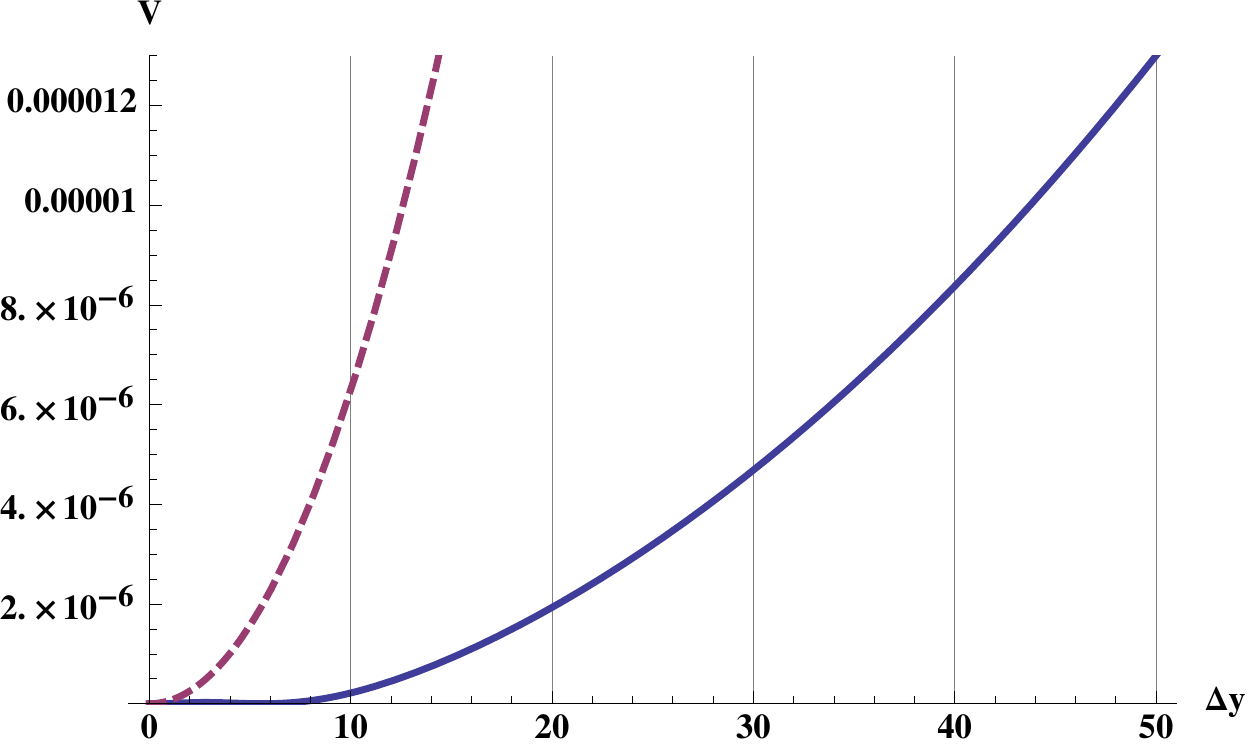}
 }
\caption{Plots of the effective inflaton potential (blue, solid) and the `naive' inflaton potential (red, dashed) vs.~$\Delta y$.}
\label{fig:n2Va}
\end{figure}

Given this numerical input, we determine the backreaction on the moduli $x$, $v^i$ and $w^i$ and calculate the effective inflaton potential as described in \ref{sec:branalytical} and \ref{sec:backreaction}. We find the following: The displacements $\delta x$, $\delta v^i$ and $\delta w^i$ are shown in figure \ref{fig:n2xa}.
One observation is that for intermediate $\Delta y \lesssim 10$ the displacements show a non-trivial dependence on $\Delta y$. However, for large $\Delta y \gtrsim 10$ they exhibit the behaviour predicted in the previous section: all $\delta v^i$ and $\delta w^i$ asymptote to a (small) constant value while $|\delta x|$ grows linearly with $\Delta y$ (with a small slope). The asymptotic behaviour for large $\Delta y$ can be quantified as
\begin{align}
\nn \delta x \rightarrow (0.229 \ \Delta y + 13.725) \cdot 10^{-4} \ , \qquad
\delta v^1 \ & \rightarrow -2.583 \cdot 10^{-4} \ , \quad \delta w^1 \rightarrow -2.700 \cdot 10^{-4} \ , \\
\nn \delta v^2 \ & \rightarrow -2.512 \cdot 10^{-4} \ , \quad \delta w^2 \rightarrow -2.268 \cdot 10^{-4} \ .
\end{align}
Also notice that the asymptotic values for $\delta v^i$ and $\delta w^i$ as well as the slope and offset in $\delta x$ are not larger than $\sim 10^{-4}$, which is the size of our small parameters $|\eta_I|$. Thus the displacements in $x$, $v^i$ and $w^i$ are small to the extent that $|\eta_I|$ are small. This is consistent with the analytic results in the previous section. 

Having analysed the backreaction on moduli, we now turn our attention to the effective inflaton potential. The result is plotted in figure \ref{fig:n2Va}. The effective inflaton potential with backreaction is displayed as a solid blue line. We also show the `naive' inflaton potential (red dashed line). While the `naive' inflaton potential is an exact parabola of the form $6.29 \cdot 10^{-8} \ (\Delta y)^2$, the effective inflation potential shows a more subtle behaviour. Most importantly, it is obvious that the effects of backreaction are by no means negligible: the naive inflaton potential is modified considerably. In particular, for intermediate $\Delta y \lesssim 10$ the behaviour of the effective inflaton potential departs from that of a simple monomial. In fact, we find an additional minimum at $\Delta y \simeq 5.64$. However, for large $\Delta y \gtrsim 10$ it is dominated by a quadratic term: 
\be
\Delta y \gtrsim 10 \ : \qquad V \simeq \ 4.73 \cdot 10^{-9} \ (\Delta y)^2 \ .
\ee 
The upshot is the following: the effective potential offers a large region of field space $\Delta y \gtrsim 10$ where quadratic inflation can be in principle realised. It is important to note that the quadratic behaviour does not persist all the way to the minimum at $\Delta y =0$. 

It also behoves to check that our findings are robust once higher order terms in the expansion of $V$ (in $\delta x$, $\delta v^i$ and $\delta w^i$) are taken into account. We can do so by adding terms of the form $\mathcal{O}(1) (\delta v^i)^3$ etc.~to our effective potential and check to what extent $V$ is affected. For large $\Delta y$ cubic corrections of the form $\mathcal{O}(1) (\delta x)^3$ are the most dangerous. One can check that cubic corrections of the form $\mathcal{O}(1) (\delta x)^3$ are strictly subleading in the interval of interest $10 < \Delta y < 100 \sim 1 / \epsilon$. Interestingly, we also find that higher order corrections will not change $V$ significantly at small and intermediate $\Delta y < 10$. While higher order corrections will modify the potential at the very bottom of the minimum at $\Delta y \approx 5.64$, corrections of the form $\mathcal{O}(1) (\delta v^i)^3$ etc.~are not large enough to destroy the existence of this additional minimum.

Nevertheless, the main observation is that for $10 \lesssim \Delta y \lesssim 100$ the effective potential is under control and essentially quadratic.

\subsection{Four-moduli-model}
\label{sec:num2mod}
Here we also present an example with inflaton modulus $u$ and three further moduli $z^1$, $z^2$ and $z^3$. The numerical values used for this example are collected in appendix \ref{app:num2}. Overall, quantities that are not required to be small are chosen to be $\mathcal{O}(1)$. Any quantities which need to be tuned small are assigned values $\mathcal{O}(10^{-4})$. Just like the previous example, this is a numerical realisation of the analysis performed in sections \ref{sec:branalytical} and \ref{sec:backreaction}.

\begin{figure}[t]
 \subfloat[][]{
 \includegraphics[width=0.46\textwidth]{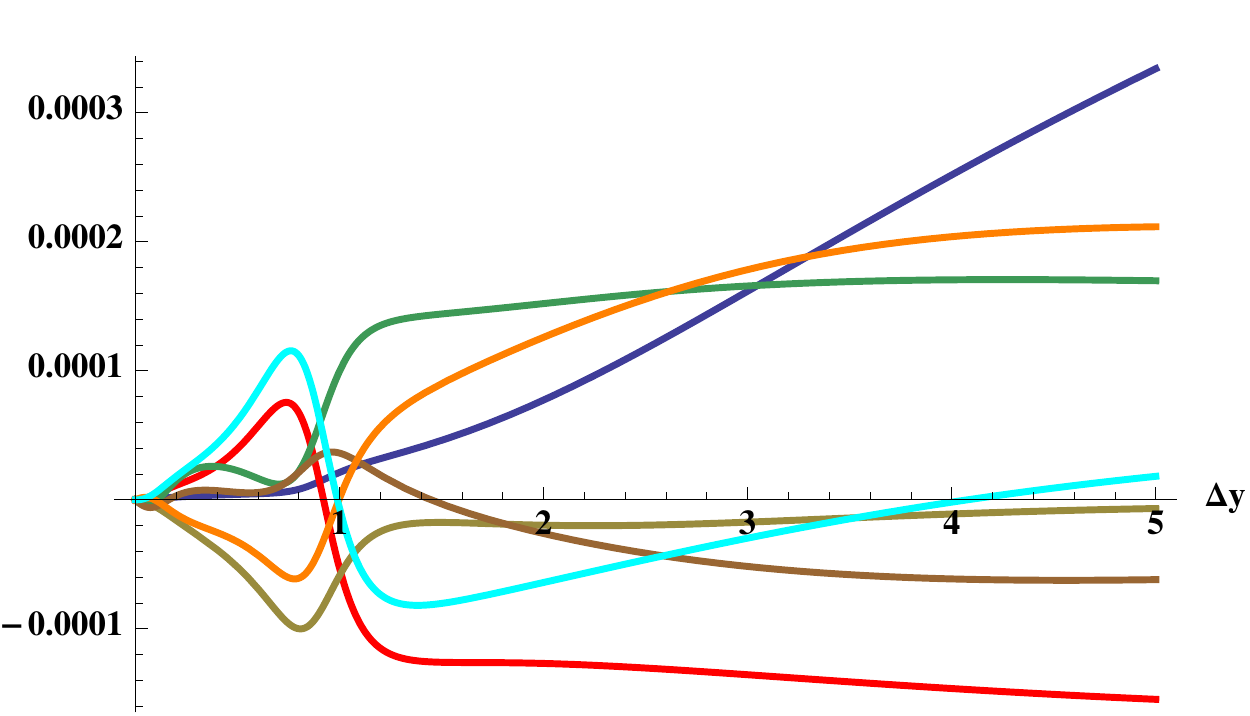}
 }
 \ \hspace{0.5mm} \hspace{5mm} \
 \subfloat[][]{
 \includegraphics[width=0.46\textwidth]{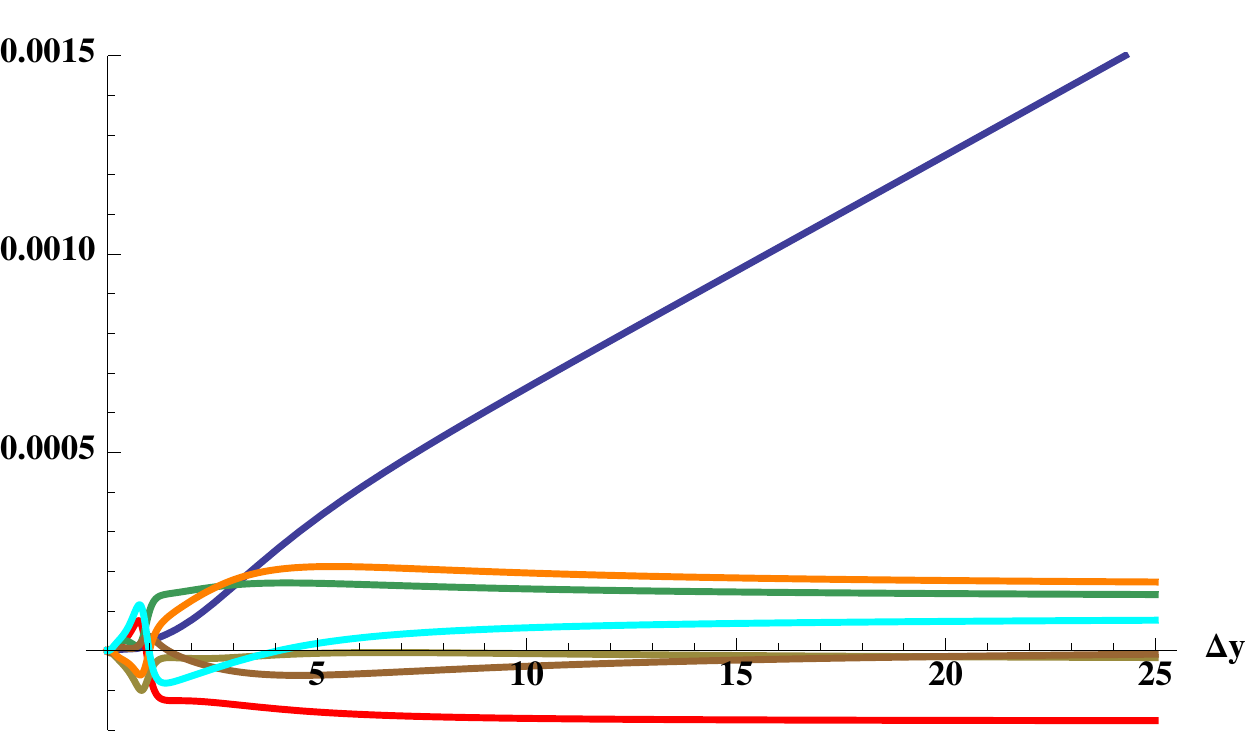}
 }
\caption{Plots of the displacements $\delta x$ (blue), $\delta v^1$ (red), $\delta w^1$ (ochre), $\delta v^2$ (green), $\delta w^2$ (brown), $\delta v^3$ (orange) and $\delta w^3$ (cyan) vs.~$\Delta y$.}
\label{fig:n4x}
\end{figure}

\begin{figure}[t]
 \subfloat[][]{
 \includegraphics[width=0.46\textwidth]{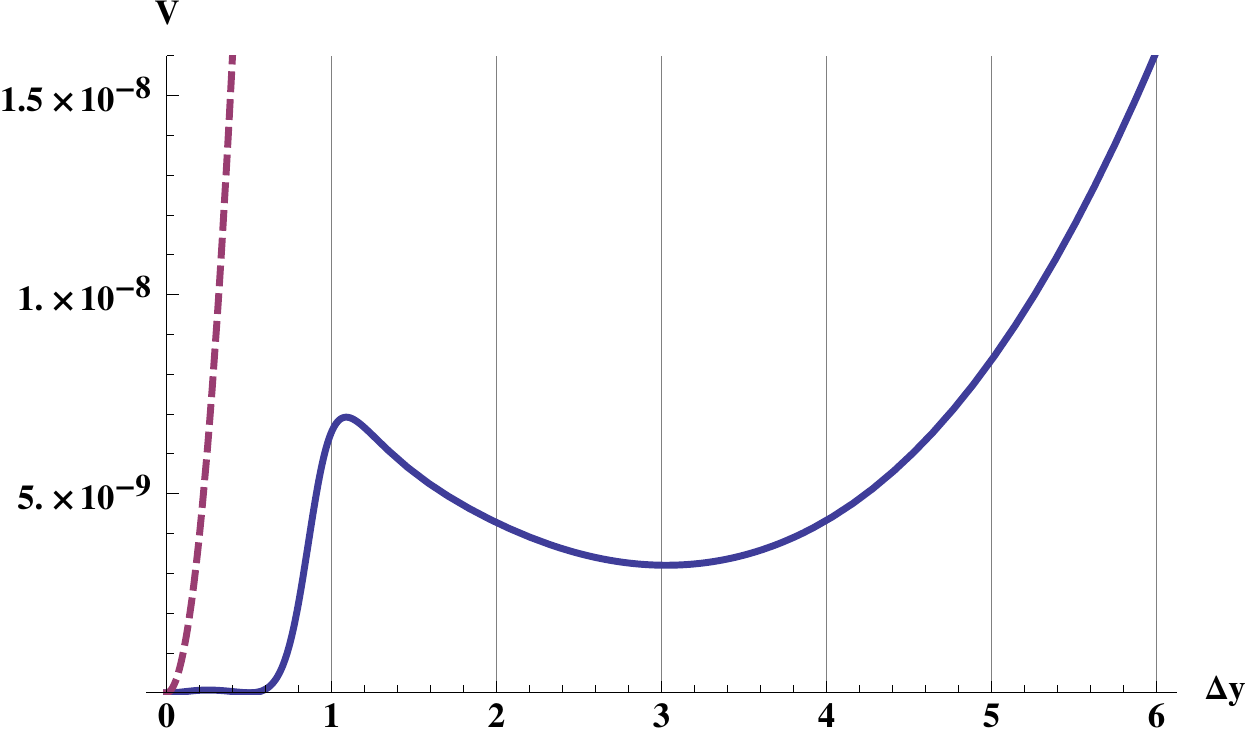}
 }
 \ \hspace{0.5mm} \hspace{5mm} \
 \subfloat[][]{
 \includegraphics[width=0.46\textwidth]{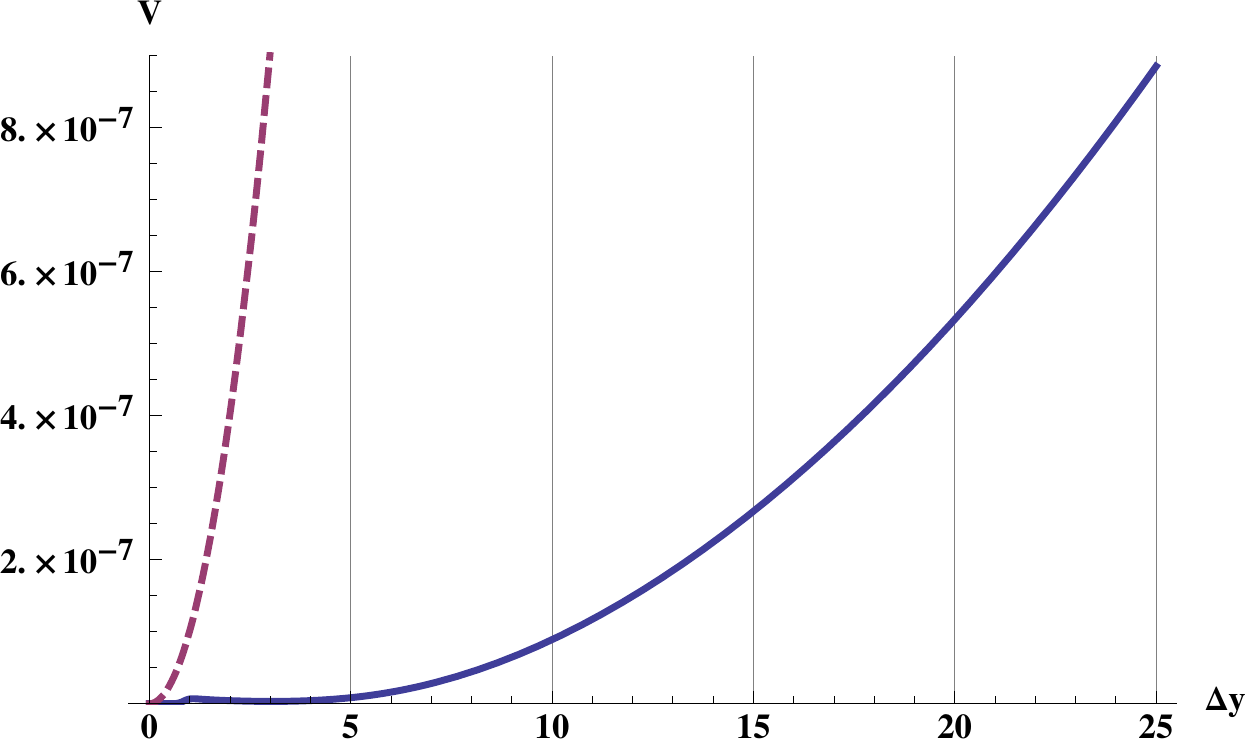}
 }
\caption{Plots of the effective inflaton potential (blue, solid) and the `naive' inflaton potential (red, dashed) vs.~$\Delta y$.}
\label{fig:n4V}
\end{figure}

We immediately proceed to the results for the displacements $\delta x$, $\delta v^1$, $\delta w^1$, $\delta v^2$, $\delta w^2$, $\delta v^3$ and $\delta w^3$, which we display in figure \ref{fig:n4x}. We again find that for large $\Delta y \gtrsim 5$ the displacements $\delta v^1$, $\delta w^1$, $\delta v^2$ and $\delta w^2$ approach a small constant value of order $\sim 10^{-4}$. Also, $\delta x$ asymptotes towards a linear function of $\Delta y$ with slope and offset of order $\sim 10^{-4}$.

The result for the effective inflaton potential (figure \ref{fig:n4V}) exhibits the expected behavior for large $\Delta y$: For $\Delta y \gtrsim 5$ the potential approximates a parabola of the form $1.65 \cdot 10^{-9} (\Delta y)^2$. However, we find an interesting behaviour for intermediate  $\Delta y$: the potential exhibits a local minimum with non-zero $V$ for $\Delta y \approx 3$. By adding terms of the form $\O(1) (\delta x)^3$ etc.~to $V$ we can also check explicitly that in the region of interest ($\Delta y < 100$) higher order terms in the expansion of $V$ can be ignored. Interestingly we find that higher order terms do not destroy the local minimum at $\Delta y \simeq 3$.

We conclude that inflation could in principle be realised in this model. For $\Delta y \gtrsim 5$ the potential is essentially quadratic and can support chaotic inflation. The inflaton would roll down the potential until it reached the local minimum at $\Delta y \simeq 3$ where inflation would end. 

We can now make an interesting observation based on the fact that the local minimum has a positive vacuum energy. Recall that K\"ahler moduli stabilisation following the Large Volume Scenario leads to an AdS minimum, which needs to be uplifted to give a dS vacuum. If our analysis in this paper can be successfully combined with K\"ahler moduli stabilisation \`a la LVS, the positive vacuum energy of the local minimum could provide the necessary uplift. Our vacuum would then be identified with the minimum we observe at $\Delta y \simeq 3$. We take this finding as a hint that the sector of complex structure moduli as studied in this paper can in principle give rise to metastable dS vacua.


\section{Tuning in the landscape}
\label{sec:landscape}
  The previous analyses of backreaction have shown the necessity of tuning of certain parameters, namely $a(z)$ and $a_i(z)$ with $i$ running over all complex structure moduli entering $a$. In sections \ref{sec:branalytical} and \ref{sec:tuneless} we found that backreaction of complex structure moduli can be controlled when we tune parameters as follows:
\begin{align}
\nn \textrm{Sec.~2.4:} \quad & |a| = \epsilon \ll 1\, \qquad |a_i+\K_i a| \sim \epsilon^2 \ , \\
\nn \textrm{Sec.~2.6:} \quad & |a| = \epsilon \ll 1\, \qquad |a_i+\K_i a| \sim \epsilon \ .
\end{align}
Let $J_t/2-1$ be the number of complex structure moduli which $a$ depends on, i.e.~$i=1,\dots,J_t/2-1$. Then $J_t$ counts the required number of tunings in both cases (note that the tuning of one complex parameter results into two tuning conditions for real parameters). In this section we provide an estimate of the number of remaining supersymmetric F-theory flux vacua after imposing the tuning conditions. In particular, we wish to count the number of vacua where $|a|$ and $|a_i + \K_i a|$ are sufficiently small, i.e.~$|a| < \epsilon$ and $|a_i+\K_i a| < \epsilon^2, \epsilon$ for setups following sections \ref{sec:branalytical} and \ref{sec:tuneless} respectively.

We will closely follow \cite{0404116} although the authors counted the number of susy flux vacua in the type IIB theory on a CY threefold $Y$, where $X = (T^2 \times Y)/\mathbb{Z}_2$.\footnote{Notice however some minor differences in the notation. While derivatives with respect to the axio-dilaton are denoted by $\partial_0$ or $D_0$ in \cite{0404116}, we write $\partial_S$ or $D_S$, respectively. The index $0$ is reserved for the inflaton field.} However, due to our no-go theorem for complex structure inflation on CY threefolds, we actually do not want to consider threefolds. Nevertheless, we follow the computation in \cite{0404116} and modify it appropriately in order to find the parametric dependence of the number of vacua on the tuning parameter $\epsilon$, also in the fourfold case.\footnote{A more precise analysis would presumably be possible using the techniques of \cite{14086156,14086167}, where the counting of vacua on fourfolds was discussed in the context of F-theory GUTs.}   

Recall that in \cite{0404116} the number of supersymmetric flux vacua satisfying the tadpole condition $L \leq L_{\star} \equiv \chi(X)/24$ on a CY fourfold $X$ was estimated to be  
\begin{equation} \label{counting formula 1}
\mathcal{N}(L \leq L_{\star}) = \frac{(2\pi L_{\star})^{2m}}{(2m)!\sqrt{\det \eta}} \int_{\mathcal{M}}d^{2m} z \ \det (g) \ \rho( z),
\end{equation}
where $\eta$ is the intersection form on $X$ and $m = h^{2,1}_-(Y)+1$. $\mathcal{M}$ denotes the moduli space over which the density $\rho$ of supersymmetric vacua (per unit volume of $\mathcal{M}$) is integrated. The authors arrived at this result by changing variables from the flux vector (of F-theory) to a set of variables $(X, Y, Z, \overline{X}, \overline{Y}, \overline{Z})$ defined by 
\begin{equation}
X \equiv \int_X G_4 \wedge \Omega_4 = W, \hspace{1cm} Y_A \equiv D_AW, \hspace{1cm} Z_I \equiv D_SD_I W
\end{equation}
in the orientifold limit. Using these variables, one can express $\rho$ as follows:
\begin{equation}
\rho(z)= \pi^{-2m}\int d^2X d^{2m-2}Z e^{-\left|X\right|^2-\left|Z\right|^2}\left|X\right|^2 \left| \det \begin{pmatrix} \delta_{IJ}\overline{X}-\frac{Z_I\overline{Z}_J}{X} & \mathcal{F}_{IJK}\overline{Z}^K \\ \overline{\mathcal{F}}_{IJK}Z^K & \delta_{IJ}X - \frac{\overline{Z}_IZ_J}{\overline{X}} \end{pmatrix} \right| \ .
\end{equation}
The tensor $\mathcal{F}_{IJK}$ has a purely geometric meaning and is defined by 
\begin{equation}
\mathcal{F}_{IJK} = \int_Y D_ID_JD_K \Omega_4 \wedge \overline{\Omega}_4 \ .
\end{equation}
The prefactor in \eqref{counting formula 1} will be modified if we impose the  $J_t$ tuning conditions. For the setup discussed in section \ref{sec:tuneless} we require $\left|a_I\right| \lesssim \epsilon$ with $a_0 \equiv a$ and $a_i = \partial_i a$ for $i=1,\dots,J_t/2-1$ and $\epsilon \ll 1$. From the Gukov-Vafa-Witten potential it is clear that the $a_I$ are linear functions of the F-theory flux vector components $N_{\alpha}$ with $\alpha=1,\dots, K=4m-J_f$, where $J_f$ counts the number of flux components chosen to be zero in order to construct a superpotential linear in $u$. 

To see how $\mathcal{N} \equiv \mathcal{N}(L \leq L_{\star}, |a_I| \lesssim \epsilon)$ differs from $\mathcal{N}(L \leq L_{\star})$ shown in \eqref{counting formula 1}, we redo the derivation in \cite{0404116} and implement the tuning conditions by including factors $\Theta(\epsilon - |a_I|)$ for all $I$ as follows:
\begin{align} \label{eq:contour}
\mathcal{N}&= \frac{1}{2\pi i} \int_C \frac{d \alpha}{\alpha}e^{\alpha L_{\star}}\mathcal{N}(\alpha), \\
\mathcal{N}(\alpha) & \simeq \int_{\mathcal{M}}d^{2m} z \int d^KNe^{-\frac{\alpha}{2}N\eta N}\delta^{2m}(DW)\left|\det D^2W \right|  \times \prod_{I=0}^{J_t/2-1}  \Theta \left(\epsilon - \left|\tilde{a}_{I\alpha}N_{\alpha}\right|\right),
\end{align}
where the $\tilde{a}_{I\alpha}$ are the coefficients of the linear expansion of $a_I$ in terms of the components of $N$, i.e.~$a_I = \tilde{a}_{I\alpha}N_{\alpha}$. The curve $C$ goes along the imaginary axis and passes the pole to the right.  It is easy to show that this gives rise to a parametric behaviour $\mathcal{N}(\alpha) \sim \alpha^{-(K-J_t)/2}$. Indeed, one can write 
\begin{align} \nonumber
& \mathcal{N}(\alpha)  \simeq \int_{\mathcal{M}}d^{2m} z \int d^{K}N  e^{-\frac{\alpha}{2}N\eta N}\delta^{2m}(DW)\left|\det D^2W \right|   \prod_{I=0}^{J_t/2-1}   \Theta \left(\epsilon - \left|\tilde{a}_{I\alpha}N_{\alpha}\right|\right) \\ 
\nonumber & = \int_{\mathcal{M}}d^{2m} z \int d^{K}\tilde{N} \alpha^{-K/2} e^{-\frac{1}{2}\tilde{N}\eta \tilde{N}}\delta^{2m}(DW)\left|\det D^2W \right| \prod_{I=0}^{J_t/2-1}  \Theta \left(\sqrt{\alpha}\epsilon - \left|\tilde{a}_{I\alpha}\tilde{N}_{\alpha}\right|\right), 
\end{align}
where we substituted $N=\tilde{N}/\sqrt{\alpha}$ and simultaneously rescaled the argument in the $\Theta$-function by $\sqrt{\alpha}$. This rescaling clearly does not modify the result but it allows to read off the parametric dependence of $\mathcal{N}$ on $\alpha$ easily: we will justify in the steps from \eqref{e1}-\eqref{e2} that the $J_t/2$ $\Theta$-factors give rise to an overall factor $\sim (\alpha \epsilon^2)^{J_t/2}$.  Hence, the parametric dependence on $\alpha$ is indeed 
\begin{equation*}
\mathcal{N}(\alpha) \simeq \alpha^{-(K-J_t)/2} \mathcal{N}(\alpha=1).
\end{equation*}
Note that without the rescaling of the argument of the $\Theta$-functions the tuning conditions would have introduced factors of $\alpha$ into the terms $\delta^{2m}(DW)\left|\det D^2W \right|$, which makes it more difficult to find the overall parametric dependence on $\alpha$.  
 Now, the contour integral \eqref{eq:contour} can be readily evaluated:
\begin{equation}
 \mathcal{N}= \frac{1}{2\pi i} \int_C \frac{d \alpha}{\alpha^{1+(K-J_t)/2}}e^{\alpha L_{\star}}\mathcal{N}(\alpha=1) = \frac{L_{\star}^{2m-(J_f+J_t)/2}}{(2m-(J_f+J_t)/2)!}\mathcal{N}(\alpha=1).
\end{equation}
Consequently, the tuning conditions modify (\ref{counting formula 1}) as follows: 
\begin{align} \label{counting formula 2}
\nn & \mathcal{N} (L \leq L_{\star}, |a_I| \lesssim \epsilon) \\
\nn \simeq \ & \frac{(2\pi)^{2m-J_f/2}L_{\star}^{2m-(J_f+J_t)/2}}{(2m-(J_f+J_t)/2)!\sqrt{\det{\eta}}} \int_{\mathcal{M}}d^{2m} z \det g \ \times \\
\nonumber & \times \pi^{-2m-J_f/2}\int d^2X d^{2m-2-J_f/2}Z e^{-\left|X\right|^2-\left|Z\right|^2}\left|X\right|^2 \left| \det \begin{pmatrix} \delta_{IJ}\overline{X}-\frac{Z_I\overline{Z}_J}{X} & \mathcal{F}_{IJK}\overline{Z}^K \\ \overline{\mathcal{F}}_{IJK}Z^K & \delta_{IJ}X - \frac{\overline{Z}_IZ_J}{\overline{X}} \end{pmatrix} \right| \times \\ 
& \times \prod_{i=0}^{J_t/2-1}  \Theta (\epsilon - |a_I(X,Z, z)|)
\end{align}
Now one can make the following change of variables: we can express some $X,Z, z$ by $a_I$, which introduces a factor $\left|\det\left(\frac{\partial(a_0,a_1,\dots,a_{J_t/2-1})}{\partial(y_0,y_1,\dots,y_{J_t/2-1})}\right)\right|$ with $\{y_0,\dots,y_{J_t/2-1}\}$ being a subset of all the $X,Z$ and $z$. We expect it to be neither particularly large nor small, since the components of the Jacobian are typically $\mathcal{O}(1)$. As a result, the number of remaining supersymmetric flux vacua is estimated to be 
\begin{equation} \label{counting formula 3}
\mathcal{N}(L \leq L_{\star}, |a_I| \lesssim \epsilon) \sim  \frac{(2\pi)^{2m-(J_f+J_t)/2}L_{\star}^{2m-(J_f+J_t)/2}}{(2m-(J_f+J_t)/2)!} \cdot (\pi \epsilon^2)^{J_t/2},
\end{equation}
where we also neglected $\sqrt{\det{\eta}}$. The factor $(2\pi)^{-J_t/2}$ arises from integrating out the tuning conditions.
The factor $ (\pi \epsilon^2)^{J_t/2}$ can be understood by the following considerations:

First of all, we can rewrite the integral in eq.~(\ref{counting formula 2}) symbolically as 
\begin{equation} \label{e1}
\mathcal{N} \sim \int_{\mathcal{M}\times \mathbb{R}^{K/2}}d^{d}x f(\vec{x})\prod_{I=0}^{J_t/2-1}  \Theta (\epsilon - |a_I(\vec{x})|),
\end{equation}
where the components of $\vec{x} \equiv (x^1,\dots, x^{d})$ with $d=2m+K/2$ replace the variables $z^I, Z_I$ and $X$. 
We assume that the combined zero locus of the $a_I$ is a $(d-J_t)$-dimensional submanifold $\mathcal{R} \subset \mathcal{M}\times \mathbb{R}^{K/2}$. Without loss of generality we parametrize this submanifold by $x^1,\dots, x^{k}$, $k=d-J_t$. The remaining variables $x^{k+1},\dots, x^{d}$ are traded for $J_t/2$ pairs of variables $\alpha_I, \beta_I$, such that $a_I = \alpha_I + i \beta_I$. Thus, we have
\begin{equation}
\mathcal{N} \sim \int_{\mathcal{M}\times \mathbb{R}^{K/2}}d^{k}x d\alpha_0d\beta_0 \dots d\alpha_{J_t/2-1}d\beta_{J_t/2-1}\tilde{f}(x^1,\dots,x^k,\vec{\alpha},\vec{\beta})\prod_{I=0}^{J_t/2-1}  \Theta (\epsilon - \sqrt{\alpha_I^2 + \beta_I^2}),
\end{equation} 
where the determinant of the Jacobian for the transformation is absorbed into $\tilde{f}$. Next, it is convenient to introduce polar coordinates $(r_I, \phi_I)$ for every pair $\alpha_I, \beta_I$. Hence, one has to evaluate 
\begin{align} \label{e2}
\mathcal{N} &\sim \int_{\mathcal{R}}d^kx \prod_{I=0}^{J_t/2-1} \int_0^{\infty} dr_I \int_{0}^{2\pi} d\phi_I r_I \Theta(\epsilon - r_I)\tilde{f}(x^1,\dots, x^k, \vec{r}, \vec{\phi}) = \\ \nonumber
& =  \int_{\mathcal{R}}d^kx  \prod_{I=0}^{J_t/2-1} \int_{0}^{\epsilon} dr_I \int_{0}^{2\pi} d\phi_I r_I \tilde{f}(x^1,\dots, x^k, \vec{r}, \vec{\phi}) \sim \left( \pi \epsilon^2 \right)^{J_t/2}, 
\end{align}
where we assumed that $\tilde{f}$ is approximately constant inside the small region of size $\sim \epsilon$. Therefore, the number of remaining flux vacua is indeed suppressed by a factor of $\sim (\pi \epsilon^2)^{J_t/2}$. 

We expect that \eqref{counting formula 3} can be used to count the remaining F-theory flux vacua by simply replacing the dimension of the flux space in type IIB by the dimension of the F-theory flux space, which is given by the Betti number $b_4$ of $X$, from which we have to subtract the number $J_f$ of flux components that had to be turned off in order to admit a linear superpotential in $u$ and in order to allow for an F-theory limit.  Thus, we use 
\begin{equation} \label{counting formula 4}
\mathcal{N}(L \leq L_{\star}, |a_I| \lesssim \epsilon) \sim  \frac{(2\pi L_{\star})^{b_4/2-(J_f+J_t)/2}}{(b_4/2-(J_f+J_t)/2)!} \cdot (\pi \epsilon^2)^{J_t/2},
\end{equation}
to estimate the number of flux vacua admitting large field inflation with complex structure moduli with tuning conditions tuning conditions $\left|a_I\right| \lesssim \epsilon$ of section \ref{sec:tuneless}.  

In section \ref{sec:branalytical} the tuning is more severe. There we have $\left|a\right| \lesssim \epsilon$ and $|D_ia|=\left|a_i+\mathcal{K}_ia\right| \lesssim \epsilon^2$. Repeating the above analysis we find that the tuning of $a$ introduces a factor of $(\pi \epsilon^2)$ into $\mathcal{N}$, while for every $|D_i a|$ which we tune small we get a contribution $(\pi \epsilon^4)$. In this case the counting formula is modified as
\begin{equation} \label{counting formula 5}
\mathcal{N}(L \leq L_{\star}, |a| \lesssim \epsilon, |D_ia| \lesssim \epsilon^2) \sim  \frac{(2\pi L_{\star})^{b_4/2-(J_f+J_t)/2}}{(b_4/2-(J_f+J_t)/2)!} \cdot \pi^{J_t/2} \epsilon^{2J_t-2}.
\end{equation}

Since the potential found in sections \ref{sec:branalytical} -- \ref{sec:tuneless} is purely quadratic for sufficiently large $\Delta y$, we can now estimate the required size of the tuning parameter $\epsilon$ for successful chaotic inflation. The inflaton potential is given by $V_{\mathrm{inf}}= \frac{1}{2}m^2 \varphi^2$, where $\varphi = \sqrt{2\mathcal{K}_{u\bar{u}}} \Delta y$ is the canonically normalised inflaton field. In order to have enough $e$-foldings (or equivalently in order to match the correct spectral index), chaotic inflation fixes the beginning of slow-roll inflation at $\varphi_{\mathrm{max}} \simeq 15$. Thus, the requirement $\Delta y < {1}/{\epsilon}$ in section \ref{sec:branalytical} turns into an upper bound for $\epsilon$:
\begin{equation} \label{upper bound on epsilon}
\epsilon < \frac{1}{\varphi \sqrt{2} \mathrm{Re}(u)} \simeq \frac{1}{15 \sqrt{2} \mathrm{Re}(u)}.
\end{equation}
Together with (\ref{counting formula 5}) we find as an upper bound for the number of supersymmetric flux vacua with the required tuning: 
\begin{equation} \label{counting formula 6}
\mathcal{N}(L \leq L_{\star}, |a| \lesssim \epsilon, |D_ia| \lesssim \epsilon^2) < \frac{(2\pi L_{\star})^{b_4/2-(J_f+J_t)/2}}{(b_4/2-(J_f+J_t)/2)!} \cdot \pi^{J_t/2}\left(\frac{1}{15 \sqrt{2} \mathrm{Re}(u)}\right)^{2J_t-2}.
\end{equation}
Unfortunately we do not know $J_f$ and $J_t$, i.e. the number of fluxes to be turned off and the number of tuning conditions. It is moreover not quite clear how large $\mathrm{Re}(u)$ can be. We therefore assume that $\mathrm{Re}(u) \sim \mathcal{O}(1)$.\footnote{Interestingly, one can derive an upper bound on $\mathrm{Re}(u)$ from the energy scale of inflation. After canonical normalisation one obtains $V_{\mathrm{inf}}\simeq \epsilon^2\varphi^2/\mathcal{V}^2 \sim 0.5 \cdot 10^{-8}$. Using \eqref{upper bound on epsilon}, one finds that $\mathrm{Re}(u)<10^4/\mathcal{V}$.} This should be sufficient to suppress the instanton corrections which scale as $\sim e^{-2\pi u}$. Nevertheless, we try to give an estimate of how large $J_t$ can be at most, assuming that $\mathrm{Re}(u) \simeq 1.2$ (equivalently $\epsilon<0.04$). Then, for the study of one particular case with $L_{\star}=972$, $b_4=23320$, $h^{3,1}(X)=3878$ (see \cite{08031194}), which gives rise to the famous number of $10^{1700}$ F-theory flux vacua, there will be a leftover of at most $\sim 10^{350}$  vacua if we require $J_t=600$ tunings (i.e. the geometry of the CY fourfold is such that only $\sim 300$ out of the $3877$ complex structure moduli (without $u$) enter $a$). In this estimate we ignored the variable $J_f$, but if it is small compared to $b_4$, this estimate should still be an appropriate approximation. Clearly, one cannot afford much more than $300$ tuning conditions due to the severe suppression factor $\epsilon^{2J_t-2}$.  However, if it is possible to realise our inflation model on a fourfold along the lines of section \ref{sec:branalytical}, such that much less than $600$ tunings are required, then there should still be a vast landscape of F-theory flux vacua left.\footnote{Furthermore, notice that for this chosen example, the integration over the flux space rather underestimates the correct value of the sum over the flux space due to the fact that the dimension of the flux space is very large. This indicates that there should be more vacua satisfying the tuning conditions left than estimated.} Note that in setups following section \ref{sec:tuneless}, where the tuning conditions are just $|a_I|\lesssim \epsilon$, the number of flux vacua is suppressed by $\epsilon^{J_t}$, see \eqref{counting formula 4}, and hence the tuning is less severe (using the above numbers, i.e. $J_t=600$ and $\epsilon = 0.04$, one has a leftover of $10^{1180}$ vacua). 

It would be interesting to work out the required tuning conditions more specifically in the future by analysing specific CY fourfolds. This would allow us to determine $J_f$ as well as $J_t$ and hence to estimate the number of remaining flux vacua more explicitly. 

Apart from the tuning conditions, the landscape will be further suppressed due to the stabilisation of $\mathrm{Re}(u)$ in the LCS limit. If, however, this requirement does not enforce too many other complex structure moduli to be stabilised in the LCS regime as well, then this constraint is not expected to be too severe. Scenarios in which all complex structure moduli are stabilised in the LCS limit are presumably difficult to realise in the string landscape.  


\section{Conclusions}

We presented a more detailed analysis of one of the recently proposed scenarios of $F$-term axion monodromy \cite{14043711}. The scenario in question is based on a complex-structure modulus $u$ which, in a partial large-complex-structure limit, features a shift-symmetric K\"ahler potential. More specifically, the imaginary part Im$(u)$ of this modulus (corresponding to the axionic part of the K\"ahler modulus of the mirror-dual type-IIA model) does not appear in the K\"ahler potential and represents a periodic variable.  Both the shift symmetry and this periodicity are then weakly broken by a flux-induced superpotential term $au$ in $W=w+au$, giving rise to a monodromy.

Making this monodromy effect weak is crucial for keeping the inflaton light, in particular lighter than the K\"ahler modulus stabilisation scale. We proposed to realise this by flux-tuning, i.e.~by a delicate cancellation of several larger contributions to the relevant, inflaton-dependent superpotential term. Thus, this superpotential term must depend on other complex-structure moduli $z$, i.e.~$au=a(z)u$, the values of which can in turn be tuned in the flux landscape.

The above raises the problem of backreaction on these other complex structure moduli $z$, which typically becomes the more severe the larger the displacement of the inflaton during inflation is. We proposed to control this issue by appealing to a further tuning: Not only does the inflaton-dependent superpotential term need to be small but, in addition, its derivatives with respect to the other complex structure moduli $z$ have to be small as well. We show that, given this additional tuning, the backreaction remains under control for a limited but potentially super-Planckian range of field displacements of the inflaton. We carefully analyse whether the substantial extra tuning which is inherent in the proposal above can indeed be realised in the landscape. Our conclusion is positive although, depending on the number of the other complex structure moduli involved in the inflaton mass term, the depletion of the number of suitable flux vacua can be severe. Thus, it is advantageous to work with geometries with many complex structure moduli of which only a small subset appears together with the inflaton in the superpotential. Searching for such concrete models would be an interesting project for the future.

While backreaction is under quantitative control, its effect on the inflaton potential is not negligible. This is the case since the original, non-backreacted potential is very flat by construction. We derive an analytic expression for the inflaton scalar potential at leading order in a set of fine-tuned, small quantities $\sim\epsilon$. However, this expression is rather complicated. It turns out that in certain regimes of the inflaton VEV, $1\ll\mbox{Im}(u)\ll 1/\epsilon$ for one way of tuning and $\mbox{Im}(u)\gg 1/\epsilon$ in another case, the expression simplifies and a purely quadratic inflationary potential can be derived. Thus, large-field (quadratic) inflation is viable and, most importantly, the potential is sufficiently flat such that K\"ahler moduli stabilisation in the Large Volume Scenario can be expected to work. It is, however, also clear that the field range cannot be made arbitrarily large if K\"ahler moduli are not to be destabilised. In particular in the second case, $\mbox{Im}(u)\gg 1/\epsilon$, the backreaction of K\"ahler moduli cannot be neglected, but inflation is still possible in principle. Since the quadratic inflationary potential derived from complex structure $F$-terms is subdominant with respect to the K\" ahler moduli backreaction, the phenomenology is more complicated in this case. 

Even more interestingly, also at small field displacements, Im$(u)\lesssim 1$, the potential becomes more complicated. On the one hand, this affects both reheating as well as the detailed calculation of inflationary predictions (the interrelation of spectral index, number of e-foldings and inflationary scale is more complicated than for simple quadratic inflation). On the other hand, the more complicated form of the inflaton potential at small VEVs raises the hope that the inflaton might not roll down all the way to the supersymmetric point $u=u_{\star}$. Instead, for some of the models of the presented class, it may get caught in a SUSY-breaking local minimum, leading to a novel $F$-term uplifting mechanism (possibly related in spirit to \cite{14064866}). The parametric smallness of this uplift would derive from the same tuning that makes the inflaton light.

Both the complicated general form of the inflaton potential as well as the special features at relatively small VEVs clearly deserve further investigation. We attempted to support our analysis by a detailed numerical investigation based on randomly generated values for the various flux-dependent coefficients in the supergravity model. This analysis confirms the general features outlined above. Nevertheless, it is clear that much more work needs to be done, both at the 4d supergravity level as well as in terms of using concrete Calabi-Yau geometries. For instance, the role of $\alpha^{\prime}$-corrections has not yet been completely clarified. Since we discussed our proposal using both the original type IIB language as well as the mirror dual type IIA language, both types of $\alpha^{\prime}$-corrections are in principle relevant. First, the ${\cal N}=2$ level $\alpha^{\prime}$-corrections on the type IIB side (which are conjectured to also apply to the ${\cal N}=1$ situation at the orientifold point) are, of course, an intrinsic part of the LVS proposal. They are hence also implicitly used in our analysis (cf.~Sec.~\ref{sec:kaehlermoduli}). Such corrections do not depend on complex structure moduli and therefore do not directly affect our inflaton. However, it would also be important to account for $\alpha^{\prime}$-corrections arising from 7-branes. In \cite{13121376} it is shown that on CY fourfolds a certain class of F-theory $\alpha^{\prime}$-corrections does not modify the functional form of the K\"ahler potential (see also \cite{14070019}). In addition, in \cite{14043040} it was argued that a more complete analysis of large-field inflation with D7-branes requires the incorporation of higher-derivative corrections to the 4d supergravity description coming from DBI terms.\footnote{This was also discussed in much more detail in \cite{14115380}, which appeared shortly after the first version of the present paper.}

Such `DBI-induced' $\alpha^{\prime}$-corrections are an important issue deserving 
further study. At present, it is not known how these corrections are reflected in the 4d ${\cal N}=1$ supergravity description. Given that at the quantitative level the present paper relies entirely on 4d supergravity, we are thus unable to assess the importance of DBI-induced $\alpha^{\prime}$-corrections reliably in our setting. We can however hope that, as in \cite{14043040, 14115380}, these corrections will flatten the potential in a benign way, without threatening the inflationary scenario. Moreover, we expect that our landscape tuning of the DBI induced scalar potential will improve the reliability of the $\alpha^{\prime}$-expansion of the DBI action. In other words, we expect higher-order terms to be less important than in a generic situation because the energy stored in the 2-from field strength is small. However, this clearly remains to be demonstrated explicitly in the future.

Furthermore, $\alpha^{\prime}$-corrections at the ${\cal N}=2$ level on the type IIA side translate into linear contributions to the period vector \eqref{period vector threefold} in type IIB (see e.g. \cite{08041248}, based on earlier analyses in \cite{Candelas:1990rm, 9707013, 0204254}). If this was all, no further modification of our analysis would be required. Again, we do not know to which extend additional effects at the ${\cal N}=1$ level with branes will be important. 

Finally, it would also be interesting to see whether combining the choice $w\gg 1$ of \cite{14097075} with our method of tuning $a\ll 1$ leads to models with better quantitative control and less fine tuning. Important phenomenological features to be addressed include the effects of possible displacements of K\"ahler moduli (along the line of \cite{07103429}) and of the oscillatory features of the potential which will be induced by the exponentially suppressed shift symmetry breaking effects (deviations from the large-complex-structure limit). We plan to address these issues in future work.

\subsubsection*{Acknowledgements}
We are indebted to Fernando Quevedo and Michele Cicoli for raising the critical issue of backreaction of the complex structure moduli shortly after our first paper appeared. We thank Timo Weigand, Stefan Theisen, Eran Palti, Renata Kallosh, Timm Wrase, M.C.~David Marsh, Clemens Wieck, Alexander Westphal, Sebastian C.~Kraus and Dominik Neuenfeld for helpful discussions. This work was partly supported by the DFG Transregional Collaborative Research Centre TRR 33 ``The Dark Universe''. During the completion of this work, P.M. and F.R. were partially supported by DFG Graduiertenkolleg GRK 1940 ``Physics Beyond the SM''. F.R. was also partially supported by the DAAD.

\appendix

\section{Comparison with backreaction in arXiv:1409.7075}
\label{sec:comparison}
In this section we review in some more detail under which circumstances backreaction of moduli cannot in general be ignored when constructing models of $F$-term axion monodromy inflation with complex structure (CS) moduli. In particular, we wish to compare our findings to the general procedure of moduli stabilisation outlined in section 5.1 of \cite{14097075}.

In the relevant setups the superpotential can be split into a part depending on the modulus $u$ containing the inflaton, and a term containing all other CS moduli:
\be
\label{eq:BHPW} W = W_{mass}(z_i) + W_{ax} (z_i, u) \ .
\ee
The scalar potential can be written as 
\be
V = V_{mass}(z_i) + V_{mix} (z_i, u) + V_{ax} (z_i,u) \ ,
\ee
where
\begin{align}
V_{mass} \ & = e^{\mathcal{K}} {\mathcal{K}}^{I \bar{J}} D_I W_{mass} \overline{D_J W_{mass}} \ , \\
V_{mix} \ & = e^{\mathcal{K}} {\mathcal{K}}^{I \bar{J}} ( D_I W_{mass} \overline{D_J W_{ax}} +\overline{D_I W_{mass}} D_J W_{ax}) \ , \\
V_{ax} \ & = e^{\mathcal{K}} {\mathcal{K}}^{I \bar{J}} D_I W_{ax} \overline{D_J W_{ax}} \ ,
\end{align}
If $W_{ax}=0$ the moduli $z_i$ are stabilised at $D_I W_{mass}=0$. In the following we will assess to what extent the moduli $z_i$ will be destabilised if we turn on $W_{ax}$ to generate an inflaton potential. 

To simplify the discussion, let us only consider a setup with two moduli $\{z, u \}$. In addition, note that both $z$ and $u$ are complex fields, so that we are still working with four degrees of freedom. To reduce notational complexity further, we pretend that all quantities, including the fields $z$ and $u$, are real in this section. While this is not realistic, the conclusions regarding the backreaction will be the same as in a more complete analysis.

To estimate the severity of backreaction let us expand the scalar potential to second order in $\delta z$ about the minimum $z=z_0$. To get the potential at $\O(\delta z^2)$ we expand the covariant derivatives $D_I W$ as follows:
\begin{align}
D_I W_{mass} (z_0 + \delta z) \ & = 0 + \pd_z (D_I W_{mass})|_{z_0} \delta z + \pd_z^2 (D_I W_{mass})|_{z_0} (\delta z)^2 + \mathcal{O}((\delta z)^3) \ , \\
D_I W_{ax} (z_0 + \delta z, u) \ & = D_I W_{ax} (z_0, u) + \pd_z (D_I W_{ax})|_{z_0,u} \delta z + \mathcal{O}((\delta z)^2) \ .
\end{align}
Note that for general $u \neq 0$ the term $D_I W_{ax} (z_0, u)$ does not vanish. Correspondingly, the scalar potential takes the form
\be
\label{eq:Vgeneral2}
V= V_{mass}''|_{z_0} (\delta z)^2 + V_{mix}'|_{z_0,u} \delta z + V_{mix}''|_{z_0,u} (\delta z)^2 + V_{ax}|_{z_0,u} + V_{ax}'|_{z_0,u} \delta z + V_{ax}''|_{z_0,u} (\delta z)^2 \ .
\ee
Minimising the scalar potential w.r.t.~$\delta z$ we can estimate the displacement $\delta z$ due to backreaction. We obtain
\be
\label{deltazlarge} \delta z \sim \frac{V_{mix}'|_{z_0,u} + V_{ax}'|_{z_0,u}}{2(V_{mass}''|_{z_0} + V_{mix}''|_{z_0,u} + V_{ax}''|_{z_0,u})} \ .
\ee 
As long as the numerator is small or the denominator large, the displacements $\delta z$ are small and retrospectively justify our expansion of $V$.

However, so far we have not taken into account any hierarchies between $W_{ax}(z,u)$ and $W_{mass}(z)$. In particular, in order to keep the inflaton field the lightest of all CS moduli, we will need to implement $W_{ax} \ll W_{mass}$ (in an appropriate sense).

\subsubsection*{Scaling up $W_{mass}$}
For one, let us scale $W_{mass} \rightarrow \lambda^2 W_{mass}$, where we now assume $\lambda \gg 1$. Physically this can be understood as a choice of flux numbers: in particular, flux parameters entering $W_{mass}$ are chosen to be considerably larger than flux parameters contributing to $W_{ax}$. However, note that there is no tuning of fluxes at this stage, i.e.~$W_{ax}$ is not parametrically smaller than the naive expectation based on $\O(1)$ flux numbers. As a result we also have 
\begin{align}
\pd_z (D_I W_{mass})|_{z_0} \ & \rightarrow \lambda \ \pd_z (D_I W_{mass})|_{z_0} \ , \\
\pd_z^2 (D_I W_{mass})|_{z_0} \ & \rightarrow \lambda \ \pd^2_z (D_I W_{mass})|_{z_0} \ , 
\end{align}
and thus 
\be
\label{eq:Vscalelambda} V \rightarrow \lambda^2 \ V_{mass}''|_{z_0} (\delta z)^2 + \lambda \ V_{mix}'|_{z_0,u} \delta z + \lambda \ V_{mix}''|_{z_0,u} (\delta z)^2 + V_{ax}|_{z_0,u} + V_{ax}'|_{z_0,u} \delta z + V_{ax}''|_{z_0,u} (\delta z)^2 \ .
\ee
By minimising $V$ we again estimate the displacement $\delta z$. In particular, to leading order in $\lambda^{-1}$ we now have
\be
\delta z \sim \lambda^{-1} \ \frac{V_{mix}'|_{z_0,u}}{2 V_{mass}''|_{z_0}} + \mathcal{O}(\lambda^{-2}) \ .
\ee
Interestingly, the displacement is now suppressed with $\lambda$. Thus, by choosing a large enough $\lambda$ backreaction is under control in principle. This is also the conclusion found in \cite{14097075}. In addition, the modulus $u$ only appears in the numerator. Then, assuming that $W_{ax}$ grows monotonically with $u$, the displacement $\delta z$ will exhibit a similar behaviour. Substituting this solution back into \eqref{eq:Vscalelambda} we arrive at a inflaton potential which grows monotonically with $u$. In principle, such a potential is suitable for realising inflation. However, note that the effect of backreaction on the potential is not negligible. The contribution of backreaction to $V$ is not parametrically smaller than $V_{ax}$.

We hence find that by establishing a hierarchy between $W_{mass}$ and $W_{ax}$ one can control backreaction on complex structure moduli. However, if this is done by scaling up $W_{mass}$ there is a possible conflict with the stabilisation of K\"ahler moduli. Note that the inflaton mass arising from the scalar potential will scale as $m_{inf} \sim 1 / \mathcal{V}$ (due to the factor $e^{\mathcal{K}}$ in $V$), but it will be independent of $\lambda$. By increasing $\lambda$ we can then make the remaining complex structure moduli parametrically heavier than the inflaton, but we cannot make the inflaton mass small in absolute terms. One way of obtaining a sufficiently light inflaton is then to allow for a large compactification volume $\mathcal{V}$. This can be achieved if the volume is stabilised according to the Large Volume Scenario \cite{0502058}. In this scheme of moduli stabilisation, the mass of the volume modulus is $m_{\mathcal{V}} \sim |W| / \mathcal{V}^{3/2}$. Together with the LVS constraint $|W| \ll \mathcal{V}^{1/3}$, which follows from $m_{3/2} \ll m_{KK}$, this implies $m_{\mathcal{V}} \ll \mathcal{V}^{-7/6} < \mathcal{V}^{-1} \sim m_{inf}$. Thus, in spite of the option of choosing $\lambda \sim W \gg 1$, it will be challenging to reverse this hierarchy by further model building. A more promising avenue towards establishing a hierarchy between $W_{mass}$ and $W_{ax}$ is then to tune $W_{ax}$ small, which we discuss this in the following.

\subsubsection*{Tuning of the inflaton mass}
In principle, the scaling $W_{mass} \rightarrow \lambda W_{mass}$ with $\lambda \gg 1$ should be equivalent to a scaling $W_{ax} \rightarrow \lambda^{-1} W_{ax}$. However, physically, there is a huge difference. For one, this corresponds to requiring a light inflaton from the outset. Further, as flux numbers are quantised they cannot be chosen small and we cannot simply scale down $W_{ax}$. Instead, we need to tune parameters small, i.e.~we require a choice of flux numbers leading to a delicate cancellation of terms in $W_{ax}(z_0,u)$. The cancellation only holds at a point $z=z_0$ and is typically spoiled for $z \neq z_0$. In addition, if $W_{ax}(z_0,u)$ is tuned small this is not automatically true for $D_I W_{ax}|_{z_0,u}$. As the inflaton dependence enters the potential through $D_I W_{ax}|_{z_0,u}$ we will also need to tune $D_I W_{ax}|_{z_0,u}$ small (for every $I$) to obtain a sufficiently flat inflaton potential. This is the approach adopted in this paper.

We thus choose fluxes such that $D_I W_{ax}$ is small at $z=z_0$. However, as this is achieved by tuning, this will not imply that the quantity $\pd_z (D_I W_{ax})|_{z_0,u}$ is small. To be able to compare results to those of the previous analysis, we write:
\be
D_I W_{ax}|_{z_0,u} \rightarrow \lambda^{-1} \ D_I W_{ax}|_{z_0,u} \quad \textrm{for } \lambda \gg 1 \ .
\ee
Note that we only introduced the parameter $\lambda$ for convenience: it is a bookkeeping device for keeping track of terms which we tune small and does not imply a scaling. We then have:
\begin{align}
V_{mass}''|_{z_0} \ & \rightarrow \hphantom{\lambda^{-1}} \ V_{mass}''|_{z_0} \ , \\
V_{mix}'|_{z_0, u} \ & \rightarrow \lambda^{-1} \ V_{mix}'|_{z_0, u} \ , \\
V_{mix}''|_{z_0, u} \ & \rightarrow \hphantom{\lambda^{-1}} \ V_{mix}''|_{z_0, u} + \mathcal{O}(\lambda^{-1}) \ , \\
V_{ax}(z_0, u) \ & \rightarrow \lambda^{-2} \ V_{ax}(z_0, u) \ , \\
V_{ax}'|_{z_0, u} \ & \rightarrow \lambda^{-1} \ V_{ax}'|_{z_0, u} \ , \\
V_{ax}''|_{z_0, u} \ & \rightarrow \hphantom{\lambda^{-1}} \ V_{ax}''|_{z_0, u} \ .
\end{align}
We can again minimise the potential w.r.t.~$\delta z$. Here we obtain
\be
\delta z \sim \lambda^{-1} \ \frac{V_{mix}'|_{z_0,u} + V_{ax}'|_{z_0,u}}{2(V_{mass}''|_{z_0} + V_{mix}''|_{z_0,u} + V_{ax}''|_{z_0,u})} + \mathcal{O}(\lambda^{-2}) \ .
\ee
As a result, we find that the size of displacements in $\delta z$ can be controlled as long as we ensure small enough $\lambda^{-1}$. In other words, displacements are small if we tune $D_I W_{ax}|_{z_0,u}$ small. However, note that $\delta z$ as a function of $u$ is not necessarily monotonically rising due to the appearance of $u$-dependence in the denominator. Thus, when resubstituting $\delta z$ into the potential, the effective inflaton potential might not be suitable for inflation. 

This leads us to the question we wish to answer in this paper. Here we will examine the backreaction on complex structure moduli in detail for a superpotential of the form
\be
W=w(z_i) + a(z_i) u \ .
\ee
We determine the effective potential and check whether it is suitable to give rise to inflation.

\section{Numerical data for example in section \ref{sec:num2mod}}
\label{app:num2}
Here we collect the numerical data giving rise to the inflationary potential shown in section \ref{sec:num2mod}. Recall that we parameterise our expansion of $D_I W$ to first order in small quantities as
\be
D_I W = (A_{Ii} + i B_{Ii}) \delta z^i + C_{Ii} \delta \zb^i + F_I \delta x + i \eta_{I} \Delta y
\ee
Also note that upper case indices run over $I=0,1, \ldots ,n$ while lower case indices take values $i=1,2, \ldots , n$. 

In the example in section \ref{sec:num2mod} the inverse K\"ahler metric is given by
\be
K^{I \bar{J}}
=\begin{pmatrix}
  1.490 \ & \ 0.615 - 0.033 \ i \ & \ 0.385 + 0.229 \ i \\
  0.615 + 0.033 \ i \ & \ 1.292 \ & \ 0.443 + 0.042 \ i \\
  0.385 - 0.229 \ i \ & \ 0.443 - 0.042 \ i \ & \ 1.483
 \end{pmatrix} \ .
\ee
We further used
\vspace{\baselineskip}

\begin{tabular}{ l l l }
  $A_{01}=-0.860 + 0.553 \ i$ , & $A_{11}=\hphantom{-} 0.701-0.793 \ i$ , & $A_{21}=\hphantom{-} 0.990-1.040 \ i$ , \\
  $A_{02}=-0.960 - 0.589 \ i$ , & $A_{12}=\hphantom{-} 1.059+0.802 \ i$ , & $A_{22}=\hphantom{-} 0.592 -1.239 \ i$ , \\
  $B_{01}=\hphantom{-} 0$ , & $B_{11}=-0.725 + 1.193 \ i$ , & $B_{21}=\hphantom{-} 0.586+1.150 \ i$ , \\
  $B_{02}=\hphantom{-} 0$ , & $B_{12}=\hphantom{-} 1.153 + 0.653 \ i$ , & $B_{22}=-0.963 +1.146 \ i$ , \\
  $C_{01}=\hphantom{-} 0.815 + 0.618 \ i$ , & $C_{11}=\hphantom{-} 0.649 + 1.323 \ i$ , & $C_{21}=\hphantom{-} 1.224 - 0.684 \ i$ , \\
  $C_{02}=\hphantom{-} 1.166 - 0.685 \ i$ , & $C_{12}=\hphantom{-} 0.839 + 0.873 \ i$ , & $C_{22}=\hphantom{-} 0.610 - 0.736 \ i$ , \\
  $F_{0 \hphantom{1}}=\hphantom{-} 1.244 - 0.997 \ i$ , & $F_{1 \hphantom{1}}=-0.731+0.880 \ i$ , & $F_{2 \hphantom{1}}=-0.769-1.490 \ i$ , \\
  $\eta_{0 \hphantom{A}}=\hphantom{-} 0.005 - 0.013 \ i$ , & $\eta_{1 \hphantom{A}}=-0.006-0.011 \ i$ , & $\eta_{2 \hphantom{A}}=\hphantom{-} 0.010 + 0.013 \ i$ . \\
\end{tabular}

\bibliography{backreactionbib}  
\bibliographystyle{JHEP}

\end{document}